\documentclass[a4paper,11pt]{scrartcl}
\usepackage{graphicx}
\usepackage{xcolor}
\usepackage{bm}
\usepackage{amsmath}
\usepackage{cite}
\usepackage{tabularx}
\usepackage{todonotes}
\usepackage{siunitx} \DeclareSIUnit\bar{bar}
\usepackage{caption}
\usepackage{subcaption}
\usepackage{soul}
\usepackage{comment}

\title{Large eddy simulations of reacting and non-reacting transcritical fuel sprays using multiphase thermodynamics}
\author{Mohamad Fathi, Stefan Hickel, and Dirk Roekaerts
\vspace{0.025in}\\
Delft University of Technology}

\date{May 13, 2022}

\begin{document}

\maketitle

\begin{abstract}
Accurate simulations of high-pressure transcritical fuel sprays are essential for the design and optimization of next-generation gas turbines, internal combustion engines, and liquid propellant rocket engines. Most important and challenging is the accurate modelling of complex real-gas effects in high-pressure environments, especially the hybrid subcritical-to-supercritical mode of evaporation during the mixing of fuel and oxidizer. In this paper, we present a novel modeling framework for high-fidelity simulations of reacting and non-reacting transcritical fuel sprays. In this method, the high-pressure jet disintegration is modeled using a diffuse interface method with multiphase thermodynamics, which combines multi-component real-fluid kinetic and caloric state equations with vapor-liquid equilibrium calculations in order to compute thermodynamic properties of the mixture at transcritical pressures. All multiphase thermodynamic formulations are presented for general cubic state equations coupled with a rapid phase-equilibrium calculation method. The proposed method represents multiphase turbulent fluid flows at transcritical pressures without relying on any semi-empirical break-up and evaporation models. Combustion source terms are evaluated using a finite-rate chemistry model, including real-gas effects based on the fugacity of the species in the mixture. The adaptive local deconvolution method (ALDM) is used as a physically consistent turbulence model for large-eddy simulation (LES). LES results show a very good agreement with available experimental data for the reacting and non-reacting ECN Spray A at transcritical operating conditions.
\end{abstract}

\section{Introduction}
In-depth understanding of turbulent reacting multiphase flows at transcritical pressures is essential for the design and optimization of efficient energy conversion systems, such as liquid rocket engines, or modern diesel engines and gas turbines. Such systems typically work based on rapid injection of cold liquid or liquid-like supercritical fuels into a chamber with an elevated pressure and temperature. Combustion occurs after transcritical evaporation during mixing of the fuel with a hot and pressurized gas or gas-like supercritical oxidizer.
% Here, \emph{transcritical} pressure refers
\emph{Transcritical} pressure refers to an operating pressure higher than the critical pressure of the pure fuel or oxidizer streams, but lower than the cricondenbar pressures of their possible mixtures. Since the cricondenbar point of hydrocarbon/air mixtures is unreachable even at elevated pressures of several hundred bars, the operating pressure of most energy conversion systems is transcritical in practice.

Whereas the operating condition of the transcritical chamber is nominally supercritical with respect to the fuel stream and a direct phase change from liquid to supercritical is expected, aligned with Gibbsian thermodynamics  and experimental reports~\cite{mayer1998atomization, singla2005transcritical, gimeno2016experimental, payri2016experimental}, the transcritical mixture can locally enter the two-phase regime and interfaces between liquid-like and gas-like phases may form. Transcritical injectors, therefore, resemble a hybrid combination of the classical two-phase disintegration with breakup and evaporation of droplets and the supercritical turbulent mixing of two dense fluids. This hybrid behavior complicates their numerical simulation; despite comprehensive experimental campaigns, it is not well understood which type of the jet disintegration dominates under exactly which conditions.

Traditional large-eddy simulation (LES) of the high-speed transcritical injection have either modeled the transcritical multiphase fluid flows by Lagrangian particle tracking (LPT) methods with sharp vapor-liquid interfaces~\cite{Pei2015, Kahila2018,Gadalla2020}, or by Eulerian single-phase dense-gas (DG) approaches with diffuse vapor-liquid interfaces~\cite{Lacaze2015,Hakim2016LES, Jofre2021}. Both LES-LPT and LES-DG approaches can be justified, but they have some inherent limitations at the transcritical conditions. The standard LPT method is very sensitive to empirical tuning parameters and was developed for the dilute mixtures, neglecting the real-fluid thermodynamics. The LES-DG approach, on the other hand, excludes the effect of the transcritical phase separation and may lead to nonphysical or ill-defined states when a part of the flow passes the meta-stable boundaries, specially at lower transcritical pressures~\cite{Ma2019}. Furthermore, some important transcritical effects such as the high solubility of the saturated liquids or the different components' evaporation rates in case of surrogate fuels are not captured by these models~\cite{Dec2000,Qiu2015,Tudisco2020}.

Using multiphase thermodynamics (MT) in the context of the diffuse-interface method has been demonstrated recently to be a promising technique to overcome the aforementioned limitations~\cite{matheis2016multi,Matheis2018}. In this formulation, the fully conservative Navier-Stokes equations (NSE) are solved for a hypothetical multi-component fluid mixture with thermo-transport properties computed using a suitable equation of state (EOS) coupled with vapor-liquid equilibrium (VLE) calculations.
Although the weak surface tension force is neglected, the method can accurately capture the physics of the problem including the subcritical region of coexisting multi-component vapor and liquid as well as the real fluid behavior such as dissolution of the ambient gas in the compressed-liquid.

Whereas LES-MT studies show excellent agreement with experimental data for the non-reacting transcritical sprays~\cite{Matheis2018, Yang2020, Koukouvinis2020}, the applicability of the method to reacting flows remained as an open question mainly due to the high computational cost of the vapor-liquid equilibrium (VLE) calculations, which increases with the number of components in the mixture, and due to the need for chemistry models that can also capture the departure from ideal-gas behavior in regions with high pressures and low temperatures. The latter can be solved using an appropriate reduced reaction mechanism~\cite{hakim2018probabilistic} and utilizing the fugacity values of species in the mixture for the evaluations of real fluid effects on the reaction rates~\cite{Kogekar2018}.

The real fluid behavior becomes apparent when using an appropriate EOS. Two-parameter cubic EOS, such as the Soave-Redlich-Kwong (SRK)~\cite{soave1972equilibrium} and Peng-Robinson (PR)~\cite{peng1976new} models, are very popular because of their computational efficiency~\cite{ma2014supercritical, matheis2016multi, Bellan2017, Wang2018, Traxinger2019}.
The intrinsic drawback of two-parameter cubic EOS is the usage of a universal critical compressibility factor, which can severely limit their accuracy close to the critical point. To overcome this limitation, volume translation methods can be employed to improve the density predictions, but the calculation of consistent caloric properties is computationally very expensive and, thus, simplified approximations are typically used in practice~\cite{Matheis2016}. Another option is utilizing a cubic EOS with three parameters. The Redlich-Kwong-Peng-Robinson (RKPR) model of Cismondi and Mollerup~\cite{cismondi2005development} calibrates the cubic EOS using the actual value of the compressibility at the critical point, and provides a consistent framework for real-fluid thermodynamic modeling at tractable computational cost~\cite{Kim2012}.

To alleviate the high computational cost of VLE calculations of reacting fluids, which typically have a very large number of components, Fathi and Hickel \cite{Fathi2021} have recently introduced a new phase-splitting method that is formulated in a reduced space and based on the molar specific values of the volume function. In comparison to the conventional method (e.g., Refs.~\cite{michelsen2004thermodynamic,Matheis2018}), the reduction method prevents the quadratic growth of the computational time with the number of components. The formulation based on the volume function leads to a better convergence behavior near the critical point and phase boundaries. In addition, the novel method directly applicable to isoenergetic‐isochoric conditions aligned with the transported variables in conservative compressible Navier-Stokes flow solvers. Combined, this yields a very considerable speed-up for the simulation of transcritical flows with many components~\cite{Fathi2021}.

In this paper, we present several novel physical and numerical models for the high-fidelity simulation of turbulent reacting and non-reacting multiphase flows at transcritical pressures. The framework is provided by a fully conservative formulation of the multi-component compressible Navier-Stokes equations, using the adaptive local deconvolution method (ALDM) for LES turbulence modelling~\cite{Hickel2014}, the RKPR EOS~\cite{cismondi2005development} coupled with the newly-introduced rapid VLE calculator~\cite{Fathi2021} for transcritical vaporization and real-fluid properties, and fugacity-based finite rate chemistry for combustion modeling.
The proposed LES-MT method is validated by comparing computational results with experimental data reported for the transcritical reacting and non-reacting Engine Combustion Network (ECN) Spray-A benchmark test cases.

\section{Physical and Numerical Models}
We use the LES-MT method for solving reacting multiphase compressible NSEs in a fully conservative form with real-fluid thermo-transport properties and fugacity-based finite-rate chemistry. In this section, the required physical and numerical models for the MT-based simulations are presented.

\subsection{Governing Equations}
The three-dimensional compressible reacting NSEs describe the conservation of mass, species, momentum, and total (absolute) internal energy:
\begin{equation}
\begin{array}{l}
\partial_{t} \rho + \boldsymbol{\nabla} \cdot(\rho \mathbf{u})=0,
\end{array}
\end{equation}
\begin{equation}
\begin{array}{l}
\partial_{t} \rho Y_{i}+\boldsymbol{\nabla} \cdot\left(\rho Y_{i} \mathbf{u}\right)=\boldsymbol{\nabla} \cdot \mathbf{j}_{i} + \dot {\omega}_{i},
\end{array}
\end{equation}
\begin{equation}
\begin{array}{l}
\partial_{t} \rho \mathbf{u}+\boldsymbol{\nabla} \cdot(\rho \mathbf{u} \mathbf{u}+p \mathbf{I} )=\boldsymbol{\nabla} \cdot \boldsymbol{\tau},
\end{array}
\end{equation}
\begin{equation}
\begin{array}{l}
\partial_{t} \rho e_t+\boldsymbol{\nabla} \cdot[(\rho e_t+p) \mathbf{u}]=\boldsymbol{\nabla} \cdot(\mathbf{u} \cdot \boldsymbol{\tau}-\mathbf{q}),
\end{array}
\end{equation}
where $\rho$ is the mixture mass density, $\mathbf{u}$ is the velocity vector, $p$ is the thermodynamic pressure of the mixture, and $e_t = e +  {\lvert \mathbf{u} \rvert}^ 2 / 2$ is the total absolute specific internal energy of the mixture and $e$ is the corresponding absolute specific internal energy. For species $i=1,2,...,N$ with $N$ being the total number of components comprising the mixture, $Y_{i}$ is the mass fraction, $\mathbf{j}_{i}$ is the diffusive mass-flux vector, and $\dot{\omega}_{i}$ is the net mass production rate. $\boldsymbol{\tau}$ and $\mathbf{q}$ denote the viscous stress tensor and the vector of heat fluxes. $\mathbf{I}$ is the unit tensor and $\boldsymbol{\nabla}$ is the gradient operator.

%\subsection{Transport properties}
The viscous stress tensor is modeled by assuming a Newtonian fluid and Stokes' hypothesis
\begin{equation}
\begin{array}{l}
\boldsymbol{\tau}=\mu\left(\boldsymbol{\nabla} \mathbf{u}+(\boldsymbol{\nabla} \mathbf{u})^{T}\right) - 2/3 \mu (\boldsymbol{\nabla} \cdot \mathbf{u})\mathbf{I}.
\end{array}
\end{equation}
The molecular viscosity coefficient $\mu$ of the multi-component mixture is estimated using Chung's correlations \cite{Poling2001}. We neglect bulk viscosity effects in our computations because accurate models to be used at pressures close to the critical point are unavailable. 

For multi-component fluid flows, the total heat flux vector 
\begin{equation}\label{eq:heatflux}
\begin{array}{l}
\mathbf{q}=-\lambda \nabla T-\sum_{i=1}^{N} { h_{i} \mathbf{j}_{i}  }
\end{array}
\end{equation}
consists of heat conduction and interspecies enthalpy diffusion, and is a function of the thermal conductivity of the mixture $\lambda$,  temperature $T$, and the partial mass absolute enthalpy $h_{i}$ of component $i$. Similar to the viscosity, the thermal conductivity of the mixture is modeled by Chung's correlations \cite{Poling2001}.

By neglecting the Soret and Dufour effects, the mass diffusion $\mathbf{j}_{i}$ is modeled using Fick's law through a simplified correlation based on the mixture averaged diffusion approximation along with an extra term to ensure zero total mass diffusion
\begin{equation}
\begin{array}{l}
\mathbf{j}_{i}=\rho D_{i} \nabla Y_{i}-Y_{i} \sum_{j=1}^{N} \rho D_{j} \nabla Y_{j}
\end{array} .
\end{equation}
The effective binary diffusion coefficient between species $i$ and the bulk mixture is approximated by
%under the assumption of equal molar diffusion velocity of all species in the bulk mixture~{\cite{turns2012introduction}} via
%
\begin{equation}\label{eq:mixdiffusivity}
\begin{array}{l}
\rho D_{i}=\left(1-X_{i}\right) / \sum_{j \neq i}^{N} X_{j}/(\rho D_{ij}),
\end{array}
\end{equation}
where $X_{i}$ is the mole fraction of species $i$, which can be computed via
$
X_{i}= Y_{i} M / M_i
$
from the mass fractions and the mean molecular weight 
$
M = 1 / ( \sum_{i}^{N} Y_{i} / M_i )
$
of the mixture, with $M_i$ being the molar mass of the species $i$.
The product of density and the binary mass diffusion coefficient of species $i$ and $j$, $\rho D_{ij}$, can be approximated accurately with Chapman and Enskog theory \cite{Poling2001} without high-pressure corrections if the system pressure is under \SI{100}{\bar}~\cite{krishna2016describing}.

\subsection{Multiphase Thermodynamics}
Simulations of multi-component fluid flows at elevated pressure require proper kinetic and caloric EOS that account for the volume of molecules and interactive forces between them in order to accurately calculate pressure and temperature from the density, internal energy, and mass composition of the mixture. The partial mass enthalpies of all components in the mixture are additionally required for the evaluation of the interspecies enthalpy diffusion flux.
At transcritical pressures, additional phase-splitting calculation should be carried out to account for the coexistence of vapor and liquid phases. In this section, we present the MT relationships required for the computations of temperature, pressure, and partial mass enthalpies of a general real fluid that can be single phase or two phases in equilibrium.

\subsubsection{Kinetic Equation of State}
A kinetic EOS represents the thermodynamic relationship between the pressure, specific volume (inverse of the density), and temperature of a single-phase substance that can be pure or a multi-component mixture.
The most popular kinetic EOS is the ideal gas (IG); however, its applicability is limited to gaseous media with a compressibility factor close to unity, that is, it only appropriate at relatively high temperatures and low pressures. For the typical pressures and temperatures of transcritical fuel sprays, cubic EOS are an attractive compromise between accuracy, model complexity, and computational cost.

The most widely used cubic EOS are SRK \cite{soave1972equilibrium} and PR \cite{peng1976new}. Both are formulated based on two model parameters and therefore have the intrinsic limitation of predicting a unique and universal compressibility factor at the critical point. This results in a systematic error for the specific volume (or density) at conditions close to the critical point. To overcome this limitation, Cismondi and Mollerup \cite{cismondi2005development} proposed three-parameter cubic EOS. This so-called RKPR EOS considers the effect of the actual critical compressibility factor as the third EOS parameter. The advantages of using the RKPR EOS for the prediction of the fluid properties at the transcritical and supercritical pressures has been highlighted previously~\cite{Martinez2006,Kim2012,Matheis2018}.

The RKPR EOS is employed for all examples discussed in the present paper, and the real-fluid thermodynamic calculations  are presented in form of the general cubic EOS
\begin{equation}\label{eq:GCEOS}
\begin{array}{l}
p ={\mathcal{R} T} / ( {\overline{\vartheta}-b})-{a}/[{\left(\overline{\vartheta}+\delta_{1} b\right)\left(\overline{\vartheta}+\delta_{2} b\right)}],
\end{array}
\end{equation}
where $\mathcal{R}$ is the universal gas constant, $\overline{\vartheta}=M/\rho$ is the molar specific volume, $a$ is the attractive energy parameter, and $b$ is the co-volume parameter. Whereas $\delta_{1}$ and $\delta_{2}$ are two constant parameters in case of two-parameter cubic EOS, they are variables in case of three-parameter cubic EOS and are computed with one extra constraint. In the RKPR EOS, the energy parameter is
\begin{equation}
\begin{array}{l}
a = ({3 \xi^{2} + 3 \xi d+ d^{2}+ d-1})/{(3 \xi+ d-1)^{2}} %\ldots \\
({\mathcal{R}^{2} T_{c}^{2}}/{p_{c}})
[3/{(2+T / T_{c})]^{m}},
\end{array}
\end{equation}
where $ d=(1+\delta_{1}^{2}) / (1+\delta_{1})$, $\xi=1+\left[2(1+\delta_{1})\right]^{\frac{1}{3}}+\left[4 / (1+\delta_{1}) \right]^{\frac{1}{3}}$, and
\begin{equation}
\begin{array}{l}
m =(-2.4407 \mathcal{Z}_c +0.0017) \omega^{2} %\ldots \\ 
+ (7.4513 \mathcal{Z}_c+1.9681) \omega + (12.5040\mathcal{Z}_c-2.7238),
\end{array}
\end{equation}
with $\mathcal{Z}_c = 1.168Z_c$ being the tuned critical compressibility factor. The co-volume parameter of the RKPR EOS is
\begin{equation}
\begin{array}{l}
b = ( {\mathcal{R} T_{c}} / {p_{c}} ) / ({3\xi+d-1}).
\end{array}
\end{equation}
The third parameter
\begin{equation}
\begin{array}{l}
\delta_{1} = 0.428+18.496 (0.338-\mathcal{Z}_c )^{0.66} +789.723 (0.338-\mathcal{Z}_c )^{2.512}
\end{array}
\end{equation}
is a function of only $\mathcal{Z}_c$ and $\delta_{2}$ follows from the constraint
\begin{equation}
\begin{array}{l}
\delta_{2} =  ( 1-\delta_{1})/( 1+\delta_{1} ).
\end{array}
\end{equation}
In these equations, $p_{c}$, $T_{c}$, $Z_{c}$, and $\omega$ are the critical pressure, critical temperature, critical compressibility factor, and acentric factor of the pure fluid.

The conventional approach to extend this pure-fluid cubic EOS to multi-component mixtures, is based on considering the mixture as a pure hypothetical substance with the parameters estimated via the van der Waals mixing rule:
\begin{equation}
\begin{array}{l}
a = \sum_{i=1}^{N} \sum_{j=1}^{N} X_{i} X_{j}{a}_{ij},
\\
b=\sum_{i=1}^{N} X_{i} b_{i},
\\
\delta_1=\sum_{i=1}^{N} X_{i} \delta_{1,i},
\\
\delta_2=\sum_{i=1}^{N} X_{i} \delta_{2,i},
\end{array}
\end{equation}
where $a_{ij}$ is obtained through a combination rule that can include any possible binary interaction effects among the components. We use the classical combination rule
\begin{equation} \label{eq:CR1}
a_{ij}=(1-\varrho_{ij}) ({a_{i} a_{j}})^{\frac{1}{2}},
\end{equation}
with $\varrho_{ij}$ being the binary interaction coefficient between components $i$ and $j$ in the mixture. In contrast to the pseudo-critical combination rule, in which $a_{ij}$ is estimated by the pure-fluid formula of the energy parameter with critical quantities estimated based on those for components $i$ and $j$, the classical combination rule \eqref{eq:CR1} provides the possibility of applying reduction methods for the phase equilibrium calculations, which can significantly reduce the computational costs for mixtures with many components in the mixture. 
Using the reduction theory~\cite{Fathi2021}, the energy parameter of the mixture can be computed through
\begin{equation}\label{eq:a-param}
\begin{array}{l}
a = \sum_{k=1}^{N_m} \lambda_{k} q_{k}^{2}  \quad \textnormal{where} \;
q_{k}\equiv\sum_{i=1}^{N} X_{i} s_{k i} {a}_{i}^{\frac{1}{2}}.
\end{array}
\end{equation}
Here, $\lambda_{k}$ and $s_{ki}$ are significant (non-zero) eigenvalues and their corresponding eigenvectors for a symmetric matrix with entries $\varrho_{ij}^{\prime}=1-\varrho_{ij}$. $N_m$ is the size of the significant eigenvalues vector. More often than not, binary interaction coefficients are set to zero due to lack of accurate data or just to simplify the computations. In that case, the matrix of $\varrho_{ij}^{\prime}$ becomes a diagonal matrix, and $N_m = 1$ regardless of the number of components in the mixture.

\paragraph{Single-Phase Pressure:} The RKPR EOS explicitly provides the thermodynamic pressure of a stable mixture whose molar composition, specific volume, and temperature are specified. The required temperature is computed via the caloric EOS, which is explained in the next subsection for non-ideal multi-component fluids.

\subsubsection{Caloric Equation of State}
A caloric EOS provides a thermodynamic relationship between a caloric property like specific internal energy, and two other properties such as temperature and specific volume for a mixture with specified molar or mass composition. Real-fluid caloric EOS can be derived by the departure function formalism using the kinetic EOS. For the general cubic EOS \eqref{eq:GCEOS}, the molar specific internal energy $\overline{e}= M e$ of a multi-component real fluid is obtained as
\begin{equation}\label{Eq:caloricEOS}
\begin{array}{l}
\overline{e}=\sum_{i=1}^{N} X_{i} \overline{h}_{i}^{\circ}(T)- \mathcal{R}T %\ldots \\
+{(a-T \partial a / \partial T )}/[ {\left(\delta_{2}-\delta_{1}\right) b}]
\ln [ ( {\overline{\vartheta}+\delta_{1} b})/({\overline{\vartheta}+\delta_{2} b} )],
\end{array}
\end{equation}
where the first two terms account for the absolute internal energy of the mixture at the actual temperature but at the standard pressure, and the last term accounts for the internal energy change via an isothermal thermodynamic path from the standard reference pressure to the actual value.

In Eq. \ref{Eq:caloricEOS}, the molar specific enthalpy at standard pressure can be computed using the NASA polynomials \cite{Burcat2005}
\begin{equation}\label{eq:nasa_h}
\begin{array}{l}
\overline{h}_i^{\circ}(T)/\mathcal{R}=-A_{i,1}/T+ A_{i,2} \ln T + \sum_{j=1}^{5} {A_{i,j+2}T^{j}}/{j}  +A_{i,8},
\end{array}
\end{equation}
where $A_{i,{1 \ldots 8}}$ are the polynomial coefficients for each component $i$, which include the formation enthalpy.

\paragraph{Single-Phase Temperature:} The temperature of a stable mixture with specified molar composition, molar specific volume, and molar specific internal energy is determined implicitly via the caloric Eq.~\ref{Eq:caloricEOS}. It can be computed numerically by an initial guess and Newton iterations:
\begin{equation}\label{eq:temperature}
T=T^*-\mathcal{L} ( \overline{e}-\overline{e}^*  ) / \overline{c}_{V}^*,
\end{equation}
where $\mathcal{L}$ is the line search parameter in order to ensure global convergence, $\overline{e}$ is the target energy, and $\overline{e}^*$ is computed via Eq.~\ref{Eq:caloricEOS} using $T^{*}$. Here, $\overline{c}_{V}^*$ is the molar specific heat capacity at constant volume and computed at temperature $T^{*}$. The thermodynamic relation required for the evaluation of $\overline{c}_{V}$ using the departure function formalism for a general cubic EOS is
\begin{equation}\label{eq:dirk_wants}
\begin{array}{l}
\overline{c}_{V}=\sum_{i=1}^{N} X_{i} \overline{c}_{P,i}^{\circ}(T)- \mathcal{R} %\ldots \\
+ {T \partial^{2} a / \partial T^{2} }[{(\delta_{1}-\delta_{2}) b}] \ln [ ( {\overline{\vartheta}+\delta_{1} b})/({\overline{\vartheta}+\delta_{2} b} )].
\end{array}
\end{equation}
The specific molar heat capacity of the component $i$ at the standard pressure is computed by taking the derivative of Eq.~\ref{eq:nasa_h} with respect to the temperature, which yields
\begin{equation}
\begin{array}{l}
\overline{c}_{P,i}^{\circ}(T)/\mathcal{R}= \sum_{j=1}^{7} {A_{i,j}T^{j-3}}.
\end{array}
\end{equation}

\subsubsection{Vapor-Liquid Equilibrium}
When vapor and liquid phases coexist, phase-splitting calculations are necessary in order to correctly determine the  thermodynamic properties of the multi-component mixture, of which overall values of molar specific internal energy $\overline{e}$, molar specific volume $\overline{\vartheta}$, and molar composition $X_i$ are known. The required phase-splitting or flash calculations are briefly explained in this section.

The two-phase equilibrium state of a fluid with $N$ components  is described by $N$ equations for species mass conservation and $2+N$ equations for temperature, pressure, and species chemical potentials of the two phases. This set of $2N+2$ equations must be supplemented with two more constraints in order to uniquely determine the molar specific volume, temperature, and molar composition of the multi-component liquid and vapor phases. These two  constraints define the type of the flash problem.
As we solve the conservative form of the governing Navier-Stokes equations, that is, transport equations for the mixture energy and density, isoenergetic-isochoric phase-splitting calculations also known as UV-flash calculations must be carried out. The two additional constraints for UV-flash calculations are 
\begin{equation}\label{eq:v_constratint}
\overline{\vartheta}=(1-\alpha)\overline{\vartheta}_l + \alpha\overline{\vartheta}_v,
\end{equation}
and
\begin{equation}
\overline{e}=(1-\alpha)\overline{e}_l + \alpha\overline{e}_v.
\end{equation}
Subscripts $l$ and $v$ refer to liquid and vapor values, and $\alpha$ is the vapor mole fraction, defined as the ratio of the mole of the vapor phase to the total mole of all phases. In order to evaluate the required specific internal energies of the liquid and vapor phases, we use Eq.~\ref{Eq:caloricEOS} for each phase separately, and to compute $\alpha$, we use total mass balance rewritten as
\begin{equation}
\begin{array}{l}
\alpha = ( M - M_l )/( M_v - M_l )
\end{array} , 
\end{equation}
with $M$ as overall average molecular weight computed using  overall mole fractions $X_{i}$. $M_l$ and $M_v$ are average molecular weights for the liquid and vapor mixtures that are computed from the liquid mole fractions $X_{l,i}$ and vapor mole fractions $X_{v,i}$, respectively. The molar compositions of liquid and vapor phases represent the solution of the phase-splitting calculations.

In order to solve the flash equations more efficiently, Fathi and Hickel~\cite{Fathi2021} recently introduced a new method that performs UV-flash calculations very fast and robust via Newton iterations with the exact Jacobian of the equilibrium temperature used for VT-flash calculations. The VT-flash here refers to isochoric-isothermal phase-splitting calculations, which this method formulates in an effectively reduced space in terms of the molar specific value of the volume function of Mikyška and Firoozabadi~\cite{mikyvska2011new} and  reduced parameters similar to those used by Nichita and Graciaa~\cite{nichita2011new}. In the following, we briefly explain how this method can be used to determine the equilibrium temperature and pressure in a general cubic EOS framework. For a comprehensive review and practical implementation guidelines, interested readers are referred to the original paper~\cite{Fathi2021}.

\paragraph{Equilibrium Temperature:}
The temperature in the two-phase region is obtained through an iterative method similar to that used for the single-phase case, but using the specific vapor and liquid internal energies and heat capacities to determine the overall values
\begin{equation}
\begin{array}{l}
\overline{e}^*=(1-\alpha)\overline{e}_{l}^* + \alpha \overline{e}_{v}^*,
\\
\overline{c}_{V}^*=(1-\alpha)\overline{c}_{V,l}^* + \alpha \overline{c}_{V,v}^*.
\end{array}
\end{equation}
Vapor and liquid quantities require the molar composition and specific volume of that phase, which are results of the rapid VT-flash calculations together with the vapor mole fraction. The iterative method can be terminated when $|\overline{e}-\overline{e}^*|/|\overline{e}| < 10^{-6}$.

According to Fathi and Hickel \cite{Fathi2021}, the VT-flash problem can be formulated based on effective reduced parameters $\overrightarrow{\mathcal{H}}=(\mathcal{H}_1^{\Delta},\ldots,\mathcal{H}_{N_m+2}^{\Delta})$ derived from Helmholtz free energy using species molar specific  volume functions. The reduced parameters are determined iteratively by applying the Newton-Raphson method for $N_m+2$ error equations
\begin{equation}
\mathcal{E}_{k} \equiv \mathcal{H}_{k,v}(\overrightarrow{\mathcal{H}}) - \mathcal{H}_{k,l}(\overrightarrow{\mathcal{H}}) - \mathcal{H}_{k}^{\Delta}=0,
\end{equation}
for $k=1,\ldots,N_m+2$ with the Jacobian matrix
\begin{equation}
\mathcal{J}_{kj} \equiv \frac{\partial \mathcal{E}_{k}}{\partial \mathcal{H}_{j}^{\Delta}}=\frac{\partial \mathcal{H}_{k,v}}{\partial \mathcal{H}_{j}^{\Delta} }-\frac{\partial \mathcal{H}_{k,l}}{\partial \mathcal{H}_{j}^{\Delta} }-\delta_{kj},
\end{equation}
for $k,j=1, \ldots,N_m+2$. In order to compute  $\overrightarrow{\mathcal{H}}_{v}$ and $\overrightarrow{\mathcal{H}}_{l}$ as well as the required partial derivatives analytically, we fist compute the K-factors 
\begin{equation} 
\begin{array}{l}
\ln \mathcal{K}_i=\sum_{k=1}^{N_m} \mathcal{H} _{k}^{\Delta}  s_{k i}+\mathcal{H} _{N_m+1}^{\Delta} b_{i}+\mathcal{H} _{N_m+2}^{\Delta}
\end{array} ,
\end{equation}
with $\mathcal{K}_i \equiv y_i/x_i$ being the K-factor of the component $i$ in the mixture.
Afterwards, the vapor mole fraction is initially obtained by the solving the classic Rachford–Rice equation
\begin{equation}
\begin{array}{l}
\sum_{i=1}^{N} {[{X}_{i}(\mathcal{K}_{i}-1)]}/{[1+\alpha (\mathcal{K}_{i}-1)]}=0.
\end{array}
\end{equation}
Next, the vapor and liquid molar compositions are obtainable from the material balances and the K-factors through
\begin{equation}
\begin{array}{l}
X_{i,l}={X}_{i}/[1+\alpha (\mathcal{K}_{i}-1 )],
\\
X_{i,v}=X_{i,l} \mathcal{K}_{i},
\end{array}
\end{equation}
for $i=1,2 \ldots N$.  From the molar compositions for both phases, we compute the required parameters of the RKPR EOS, including $a,q,b,\delta_{1},\delta_{2}$ through the relations given in the previous section for a stable single-phase mixture. The specific molar volume of the vapor is then evaluated through the isochoric constraint
\begin{equation}
\begin{array}{l}
\overline{\vartheta}_v = [\overline{\vartheta} - (1-\alpha) \overline{\vartheta}_l ] / \alpha
\end{array}.
\end{equation}
Requiring equality of pressures between two phases, the liquid specific molar volume for the general cubic EOS is the solution of a fifth-order polynomial 
\begin{equation}
\varsigma_{5} \overline{\vartheta}_l^{5}+s_{4} \overline{\vartheta}_l^{4}+\varsigma_{3} \overline{\vartheta}_l^{3}+\varsigma_{2} \overline{\vartheta}_l^{2}+\varsigma_{1} \overline{\vartheta}_l+\varsigma_{0}=0.
\end{equation}
The polynomial coefficients $\varsigma_{0..5}$ are listed in Ref.~\cite{Fathi2021}. They are functions of the parameters that we computed in the previous steps. Finally, $\mathcal{H}_{k,l}$ or $\mathcal{H}_{k,v}$ is computed from
\begin{equation}\label{eq:H1}
\begin{array}{l}
\mathcal{H}_{k}= {2 \lambda_{k} q_{k} }
{\ln \left[({\bar{\vartheta}+\delta_{1} b})/({\bar{\vartheta}+\delta_{2} b}) \right]}
/[(\delta_{1}-\delta_{2}) b\mathcal{R}T],
\end{array}
\end{equation}
for $k \leq N_m$,
\begin{equation}\label{eq:H2}
\begin{array}{l}
\mathcal{H}_{k}=
a
/ (\mathcal{R}Tb^2)
\{ { \overline{\vartheta}  b} / [( \overline{\vartheta}+\delta_{1} b )  ( \overline{\vartheta}+\delta_{2} b )] \ldots \\
- {\ln \left[({\bar{\vartheta}+\delta_{1} b})/({\bar{\vartheta}+\delta_{2} b}) \right]} / (\delta_{1}-\delta_{2})\}
-(v-b)^{-1},
\end{array}
\end{equation}
for $k = N_m+1$, and
\begin{equation}\label{eq:H3}
\begin{array}{l}
\mathcal{H}_{k}=\ln (\overline{\vartheta}-b),
\end{array}
\end{equation}
for $k = N_m+2$. Note that the subscripts $l$ and $v$ are dropped for simplicity as these equations apply to both phases.
The required equations are listed in Ref.~\cite{Fathi2021} including the analytical expression of the components of the Jacobian matrix.

Starting the procedure requires an initial guess for the vector  $\overrightarrow{\mathcal{H}}=(\mathcal{H}_1^{\Delta},\ldots,\mathcal{H}_{N_m+2}^{\Delta})$. This can be conducted using K-factor values by following the steps from the solution of the Rachford-Rice until the evaluation of $\mathcal{H}_{k,l}$ and $\mathcal{H}_{k,v}$. Then, the vector of reduced parameters can be computed using its definition via
\begin{equation}
\begin{array}{l}
\mathcal{H}_{k}^{\Delta}  = \mathcal{H}_{k,v} - \mathcal{H}_{k,l},
\end{array}
\end{equation}
for $k=1,\ldots,N_m+2$. Typically the K-factors are available from previous time steps in computational fluid dynamics simulations. In case of blind flashes where no previous equilibrium information is available, one can use Wilson's correlation as an initial guess.

\paragraph{Equilibrium Pressure}
Neglecting the surface tension force, the pressure is equal for both phases in equilibrium. As we used this equality for the determination of specific volumes, the equilibrium pressure is the saturation pressure of vapor or liquid obtained by the flash calculations at the equilibrium temperature.

\subsubsection{Partial Enthalpy}
By definition, the partial derivative of a quantity with respect to the mole fraction of a species while keeping temperature, pressure, and mole fractions of the other species unchanged is called the partial molar value of that quantity.  
It can be shown that thermodynamic relationships among extensive quantities are also valid for corresponding partial values. In this section, the calculation of the partial (mass) enthalpy in the multiphase thermodynamics framework is briefly explained, as it is required for the heat flux \eqref{eq:heatflux} and for some discretization schemes.

The (mass-basis) partial enthalpies can be computed easily from the mole-basis ones:
$
{{h}}_{i}={\overline{h}}_{i}/M_i.
$
The partial molar enthalpy for a single-phase mixture with known temperature, specific volume and molar composition can be computed via the thermodynamic relation
\begin{equation}
{\overline{h}}_{i}={\overline{e}}_{i}+p\overline{\vartheta}_{i},
\end{equation}
where $\overline{\vartheta}_{i} \equiv ( \partial \overline{\vartheta} / \partial {X_k})\vert_{T,P, X_{j \neq k}} $ is the partial molar volume, which can be obtained from its definition for the considered kinetic EOS \eqref{eq:GCEOS} as 
\begin{equation}
\begin{array}{l}
{\overline{\vartheta}}_{i}=
-\{
\mathcal{R} T / (\overline{\vartheta}-b)
+ \mathcal{R} T b_{i} / (\overline{\vartheta}-b)^2
\ldots \\
+ a b_i [2b \delta_{1} \delta_{2} + \overline{\vartheta} ( \delta_{1} +  \delta_{2} ) ] / [  ( \overline{\vartheta}+\delta_{1} b )  (\overline{\vartheta}+\delta_{2} b )  ]^2
\ldots \\
-     a_i  / [ ( \overline{\vartheta}+\delta_{1} b ) ( \overline{\vartheta}+\delta_{2} b ) ]
\} / ({\partial p} / {\partial \overline{\vartheta}} )\vert_{T, X_{j}}
\
\end{array} ,
\end{equation}
where $a_i \equiv (  \partial a /  \partial X_i )\vert_{T,P, X_{j \neq i}} $  can be computed using reduced parameters \cite{Fathi2021}
\begin{equation}
\begin{array}{l}
a_i = - 2(\sum_{k=1}^{N_{m}} \lambda_{k} q_{k} s_{k i} a_{i}^{1 / 2} )
\end{array}
\end{equation}
with $i$ being species index, and
\begin{equation}
\begin{array}{l}
({\partial p} / {\partial \overline{\vartheta}})\vert_{T, X_{j}}=- {\mathcal{R} T}/{(\overline{\vartheta}-b)^{2}}
%
%\ldots \\
%
+ {a\left[2 \overline{\vartheta}+\left(\delta_{1}+\delta_{2}\right) b\right]}
/ [  ( \overline{\vartheta}+\delta_{1} b )  ( \overline{\vartheta}+\delta_{2} b )  ]^2.
\end{array}
\end{equation}
The partial molar internal energy $\overline{e}_{i}  \equiv ( \partial \overline{e} / \partial {X_k})\vert_{T,P, X_{j \neq k}} $ can be computed from the caloric EOS as 
\begin{equation}
\begin{array}{l}
{\overline{e}}_{i} =
\bar{h}_{i}^{\circ}(T)
- \mathcal{R} T
+  {( \overline{\vartheta}_{i} -  \overline{\vartheta} b_i / b )( a - T \partial a / \partial T  ) }/ [ { ( \overline{\vartheta}+\delta_{1} b ) ( \overline{\vartheta}+\delta_{2} b ) } ]
 \ldots \\  + [  T ( b_i \partial a / \partial T
-  b \partial a_i / \partial T ) + b a_i - a b_i  ] %\ldots \\
 {\ln \left[({\bar{\vartheta}+\delta_{1} b})/({\bar{\vartheta}+\delta_{2} b}) \right]}/[{\left(\delta_{2}-\delta_{1}\right) b^2}].
\end{array}
\end{equation}
Note that we have neglected the dependence of $\delta_1$ and $\delta_2$ on the molar composition in the above proposed equations following the approach used by Gmehling et al.~\cite{gmehling2019chemical} for SRK and PR EOS.

\paragraph{Equilibrium Partial Enthalpy:} If the mixture is unstable in a single phase, then phase-equilibrium calculations are necessary in order to determine the thermodynamic properties of a stable mixture of saturated vapor and liquid phases. The  overall partial enthalpy of the component $i$ can then be estimated as
\begin{equation}
\begin{array}{l}
{\overline{h}}_{i}\simeq (1-\alpha) X_{i,l} {\overline{h}}_{i,l} /  X_{i} + \alpha X_{i,v} {\overline{h}}_{i,v} / X_{i}.
\end{array}
\end{equation}

\subsection{Real Finite-Rate Chemistry}\label{sec:finite_rate}
Chemical reactions change the composition of the real-fluid mixture via the formation and destruction of species, which are expressed in the species mass balance equations by net mass production rates $\dot{\omega}_{i}$.
At elevated pressures, the evaluation of $\dot{\omega}_{i}$ requires some real-gas considerations concerning the species concentrations, which are different from the usually used ideal-gas definitions. This is addressed in this section for  finite-rate chemistry modeling.

Similar to the mass action law of the elementary reactions, the species net mass production rates of reactions with non-integer stoichiometric coefficients can also be expressed as 
\begin{equation}\label{eq:mass_rates}
\begin{array}{l}
\dot{\omega}_{i} = M_{i}
\sum_{r=1}^{N_{r}} (\nu_{ir}^\mathcal{P}-\nu_{ir}^\mathcal{R} )
\mathcal{Q}_r,
\end{array} 
\end{equation}
where $N_r$ is the total number of reactions, and $\nu_{ir}^\mathcal{P}$ and $\nu_{ir}^\mathcal{R}$ are the positive molar stoichiometric coefficients of species $i$ on the right (product) and left (reactant) side of the reaction $r$, which might be non-integer numbers in general. The reaction rate $\mathcal{Q}_r$ of the reaction $r$ depends on the concentrations of the reactants through
\begin{equation}\label{eq:reaction_rates}
\begin{array}{l}
\mathcal{Q}_r =
k_{f,r} \prod_{j=1}^{N} \mathcal{C}_{j}^{n^{\prime}_{jr}}
-k_{b,r} \prod_{j=1}^{N} \mathcal{C}_{j}^{n^{\prime \prime}_{jr}}.
\end{array}
\end{equation}
The forward and backward rate coefficients $k_{fr}$ and $k_{br}$ are are usually expressed by the Arrhenius law
\begin{equation}
\begin{array}{l}
 k(T) = \mathcal{A} \exp [-E_a / (\mathcal{R} T) ],
\end{array}
\end{equation}
as an exponential function of the temperature, with $\mathcal{A}$ being the pre-exponential factor and $E_a$ the reaction activation energy.
In addition, $n^{\prime}_{jr}$ and $n^{\prime \prime}_{jr}$ are reaction orders with respect to species $j$ in the forward and backward reactions. They might be listed separately for global reactions, otherwise they are considered equal to the molar {stoichiometric} coefficients of that species  in the reactants and products.

It is important to note that the concentrations $\mathcal{C}_j$ of the species $j$ in Eq.~\ref{eq:mass_rates} should be computed in a thermodynamically consistent way for the dense gaseous mixture. We propose to use the species' fugacity for this purpose, using the simple relation 
\begin{equation}\label{eq:conc_species}
\begin{array}{l}
\mathcal{C}_j \simeq  f_j / (\mathcal{R}T),
\end{array}
\end{equation}
with $f_j$ being the fugacity of species $j$ in the mixture. The fugacity is computed from the fugacity coefficient $\varphi_j = f_j / (X_j p)$, which can be obtained using the molar specific volume function
\begin{equation}
\begin{array}{l}
\varphi_j =   \mathcal{R}T  / ( p \psi_j ),
\end{array}
\end{equation}
with $\psi_j$ being the molar specific value of volume function for species $j$, see Ref.~\cite{Fathi2021}. Using the definition of the fugacity coefficient and the molar specific volume function, the required concentration can therefore be computed via the following simple relation for real multi-component gases
\begin{equation}
\begin{array}{l}
\mathcal{C}_j =  X_j / \psi_j
\end{array}
\end{equation}
where $\psi_j$ can be computed using the effective reduced parameters
\begin{equation} 
\begin{array}{l}
\ln \psi_j=\sum_{k=1}^{N_m} \mathcal{H}_{k}  s_{k j}+\mathcal{H}_{N_m+1} b_{j}+\mathcal{H}_{N_m+2},
\end{array}
\end{equation}
with $\mathcal{H}_{1},\ldots,\mathcal{H}_{N_m+2}$ computed using the gas or supercritical mixture temperature, composition and specific volume via  Eqs.~\ref{eq:H1}-\ref{eq:H3}. Note that in the limit of very high temperature, the fugacity coefficient tends towards unity and the ideal-gas mixture formula $ \mathcal{C}_j = X_j p / ( \mathcal{R} T) $ is recovered consistently.

Using Eq.~\ref{eq:conc_species} simplifies the computation of the backward reaction coefficient whenever it is not explicitly provided. In such cases, the backward rate coefficient should be computed using the equilibrium constant $K_{\textnormal{c}}=k_f / k_{b}$. If the reaction rates are expressed as proposed in terms of the fugacity, the equilibrium constant can be computed similarly to the ideal-gas formula in \cite{Diegelmann2016} from
\begin{equation}
\begin{array}{l}
K_{\textnormal{c},r}= [ p^{\circ} / ( \mathcal{R} T ) ]^{ \nu_{r} }
 \exp  [{\Delta S_{r}^{\circ}} / \mathcal{R}-{\Delta H_{r}^{\circ}} / ( \mathcal{R} T ) ],
 \end{array}
\end{equation}
with $\nu_{r}=\sum_{i=1}^{N}{( \nu_{ir}^\mathcal{P} - \nu_{ir}^\mathcal{R} )}$ being the net change in the number of species in the reaction, and $\Delta H_{r}^{\circ}$ and $\Delta S_{r}^{\circ}$ being the reaction enthalpy and entropy net change at the standard pressure $p^{\circ}=1$atm.

\subsection{Large Eddy Simulation}
In this work, LES is utilized for turbulence modeling. The basic idea is to focus the computational effort on the time-accurate evolution of the large scales of fluid flows that encompass almost all of the mechanical energy of the turbulent fluid motion. The evolution of the large scales is governed by a coarse-grained or low-pass filtered form of the Navier-Stokes equations. The effect of nonlinear interactions between the large resolved scales and the small truncated ones is taken into account via an appropriate modeling of the subgrid-scale (SGS) stress tensor.

In explicit LES, the SGS tensor is approximated based on the continuous filtered differential equations and model expression depending on, e.g., the filtered strain rate and possibly other quantities for which additional transport equations are needed, and the discretization of the continuous filtered transport and model equations in space and time is carried out afterwards using standard approximation theory. In implicit LES, however, the coarse-grained discrete numerical model equations are directly derived in a single approximation step that includes modeling of the effects of unresolved interactions consistent with a specifically tailored high-order numerical discretization \cite{Hickel2006,Hickel2014}. 

The LES presented in this paper have been performed with the adaptive local deconvolution method (ALDM) of Hickel et al.~\cite{Hickel2014}, which is a physically consistent implicit LES method based on a nonlinear, solution adaptive finite-volume discretization scheme and spectral turbulence theory. ALDM is appropriate for LES of the full Mach number range without any user-defined model parameters. Its superior performance in comparison to common explicit LES models has been reported by M{\"{u}}ller et al.~\cite{Muller2016} for the transcritical and supercritical injections of pure nitrogen. As discussed in \cite{Matheis2018}, ALDM does not account for all additional SGS quantities that appear after low-pass filtering the non-linear equations used for the evaluation of the thermo-transport properties and chemistry sources in transcritical flows. Although novel ideas towards modelling some of these terms have been proposed, their performance for transcritical conditions at high Reynolds number remains unclear due to a lack of experimental data for high-order statistics and spatial details. ALDM has been extensively verified and validated for transcritical injector flows; for example, by Matheis et al.~\cite{Matheis2016} for LES of Tani's cryogenic coaxial injector, by M{\"{u}}ller et al.~\cite{Mueller2016_oschwald} for LES of Oschwald's coaxial injector, and by Matheis and Hickel \cite{Matheis2018} for LES of the non-reacting ECN Spray-A.

\subsection{Numerical Implementation}
Real-fluid multiphase thermodynamics and the fugacity-based finite rate chemistry model are implemented in our fluid flow solver INCA (https://www.inca-cfd.com). We employ the same discretization schemes for the transport equations as Matheis and Hickel \cite{Matheis2018}. In this subsection, these schemes are briefly reviewed.

The governing equations are discretized in space by a conservative finite-volume scheme that uses a second-order central difference method for the diffusion terms, and ALDM for the inviscid fluxes. The van~Albada limiter \cite{VanAlbada1997} is utilized for the mass and energy flux reconstruction to avoid spurious oscillations with possible under- or overshoots at sharp density gradients.

A second-order Strang-splitting scheme is utilized to separate the stiff chemical reactions from the advection and diffusion mechanisms. The method consists of three major steps: First, the solution is updated by considering only the reaction source terms in half a time step by means of a stiff ODE solver. We used VODE \cite{Brown1989} for this purpose, which is a variable-coefficient implicit solver based on 5th-order backward differentiation formulas. Second, the obtained solution is set as initial condition for a full advection and diffusion time step with the strong stability preserving third-order explicit Runge-Kutta scheme of Gottlieb and Shu~\cite{gottlieb1998total}. Finally, the solution is updated with the second half reaction step. The time step size is adapted dynamically according to the Courant–Friedrichs–Lewy stability condition with CFL=1.

\section{Transcritical ECN Spray-A}
\subsection{Experimental Setups}
The standard reacting and non-reacting test cases of ECN Spray-A~\cite{pickett2010comparison} are selected for validation and demonstration. The fuel is pure n-dodecane ($\mathrm{C}_{12} \mathrm{H}_{26}$) with a low temperature of \SI{363}{\kelvin}, and is injected into a chamber with a pressure of \SI{60}{\bar} and a temperature of \SI{900}{\kelvin}.
At these conditions, the cold liquid n-dodecane experiences a transcritical vaporization process, since the chamber pressure is much higher than the critical pressure (\SI{18.2}{\bar}) of n-dodecane. The injection rail pressure is \SI{1500}{\bar} and the injector consists of a hydro-eroded single-hole nozzle with the nominal diameter size of $D = \SI{0.09}{\milli\meter}$ leading to the injection of about \SI{3.5}{\milli\gram} of n-dodecane fuel with a relatively constant jet velocity of about \SI{600}{\meter\per\second} during the injection time of about \SI{1.5}{\milli\second}. The initial molar compositions of the gas in the chamber are listed in Table~\ref{table:molarcompositions} for the reacting and non-reacting cases. The chamber gas is mostly nitrogen, to which oxygen is added for the reacting case. Elevated concentrations of carbon dioxide and water vapor result from burning acetylene along with hydrogen in the preparation stage.

\begin{table}[!htb]
\centering
\caption{Mole percentage of species in the chamber of reacting and non-reacting cases.}
\label{table:molarcompositions}
\begin{tabular}[]{|c|c|c|c|c|c|}
\hline
 & $\mathrm{O}_{2}$ & $\mathrm{N}_{2}$ & $\mathrm{C} \mathrm{O}_{2}$ & $\mathrm{H}_{2} \mathrm{O}$   \\
\hline
Non-reacting & 0.00 & 89.71 & 6.52 & 3.77   \\
\hline
Reacting & 15.00 & 75.15 & 6.23 & 3.62   \\
\hline
\end{tabular}
\end{table}

\subsection{Computational Setup}

\subsubsection{Grid Specifications}
\begin{figure}[!tb]
\centering
\includegraphics[height=44mm]{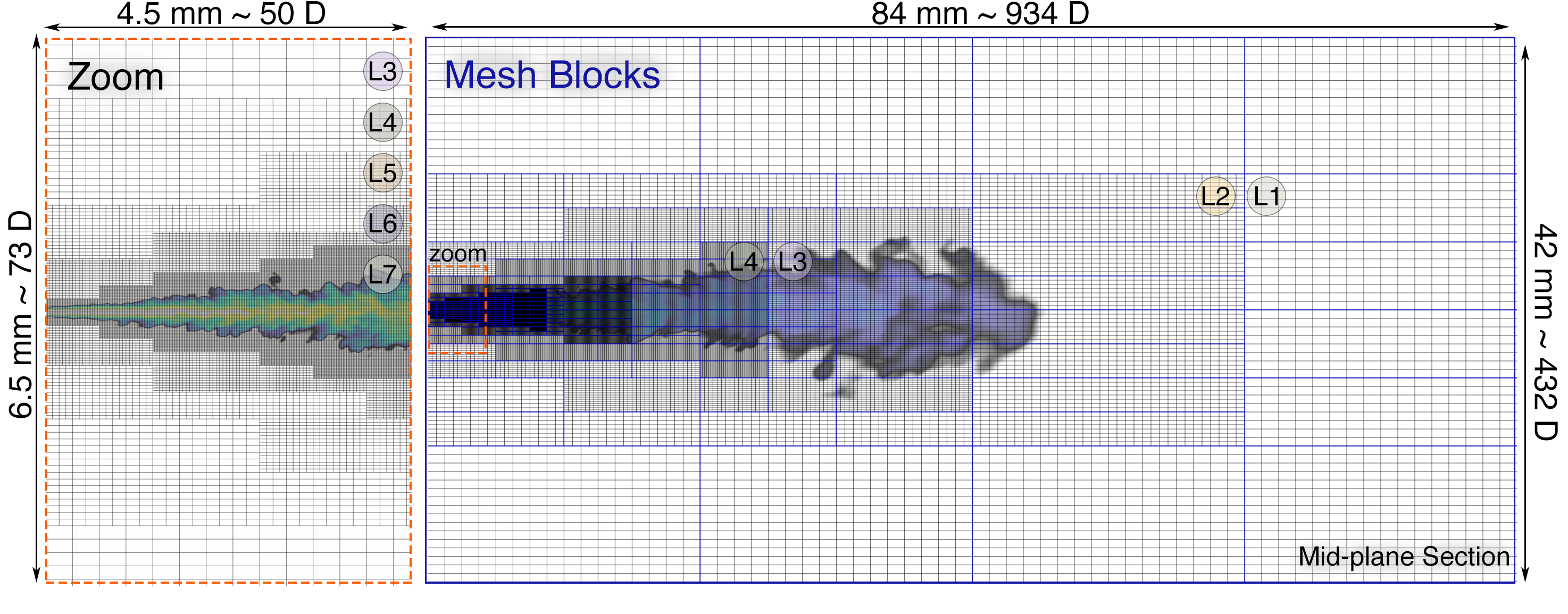}
\caption{A mid-plane section of the 3D multi-block structured grid with \si{7} levels of mesh refinements generated for LES-MT of reacting and non-reacting ECN Spray-A.}\label{fig:grid}
\end{figure}

All simulations have been carried out in a three-dimensional rectangular domain with a length of $\num{934}~D$ in the axial direction and $\num{467}~D$ in the lateral directions, with $D = \SI{0.09}{\milli\meter}$ being the nominal diameter of the injector nozzle. To minimize the total number of computational cells, we applied adaptive mesh refinement with several predefined local refinement regions towards the injector nozzle and within a hypothetical cone with spreading angle of \SI{10}{\degree}; the resulting mesh is shown in Fig.~\ref{fig:grid}. The multi-block structured grid consists of \num{2864} blocks and \num{12.7e6} cells with seven resolution levels, which are marked by L1 to L7 in Fig.~\ref{fig:grid}. Around \SI{40}{\percent} of the cells belong to the finest level with $\Delta y_{\textnormal{min}}=\Delta z_{\textnormal{min}} \simeq \SI{10.25}{\micro\meter}$ and $\Delta x_{\textnormal{min}}=2\Delta y_{\textnormal{min}}$ located in the near-nozzle region ($x/D<60$), see the zoomed area on the left side in the figure.

\subsubsection{Boundary Conditions}
\begin{figure}[!htb]
\centering
\includegraphics[height=44mm]{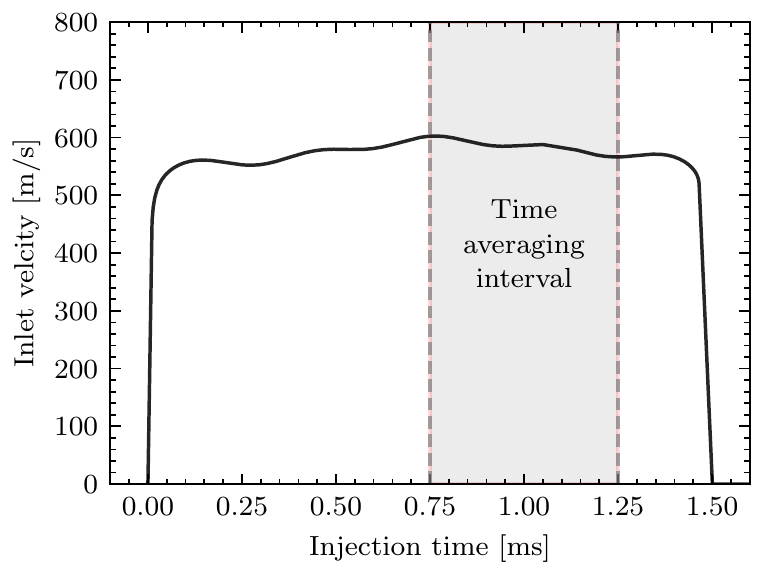}
\caption{Transient injection velocity profile used for the simulation of reacting and non-reacting ECN Spray-A test cases, computed for using RKPR EOS.}\label{fig:transientvelocity}
\end{figure}

Subsonic pressure outflow conditions are imposed at the exit face, using a constant static pressure of \SI{60}{\bar} and by extrapolating all conservative flow variables from the domain inside. 
The injector nozzle is not resolved, instead a transient velocity inflow is used at the injector's nozzle exit plane.
This inflow is pure n-dodecane at a temperature of \SI{363}{\kelvin} and a pressure of \SI{60}{\bar} with a transient axial velocity that provides the same amount of momentum as the experiment. We calculated the mass flow rate with the CMT virtual injection rate generator (http://www.cmt.upv.es) with input parameters according to the experimental conditions and a fuel density of \SI{687.24}{\kilo\gram\per\cubic\meter} consistent with the RKPR EOS at the given pressure and temperature. Fig.~\ref{fig:transientvelocity} shows the generated transient injection velocity. The transient profile is directly used as boundary condition without adding any artificial turbulent fluctuations at the boundary. The induced shear and hydrodynamic pressure fluctuations at \SI{600}{\meter\per\second} are expected to be strong enough to create turbulence almost instantaneously. Adiabatic no-slip conditions are applied for all other boundaries of the rectangular domain.

\subsubsection{Reaction Mechanism}
The finite-rate chemistry model formulated for the real-fluid effects in section~\ref{sec:finite_rate} is used for the calculation of chemical source terms in the species mass conservation equations. Due to the already high computational load of {the} LES-MT method, a {highly}-reduced reaction mechanism is selected. It is a global two-step reaction mechanism calibrated using Bayesian inference for the precise prediction of the ignition delay time of n-dodecane combustion at the experimental condition of the ECN Spray-A test case \cite{Hakim2016modeling}.
The reduced mechanism has \num{6} species.  Similar to the  approach of Westbrook and Dryer for the oxidation of paraffins, the incomplete oxidation of n-dodecane is accounted for by two-step reactions as given below:
\begin{equation}
\begin{array}{l}
\mathrm{C}_{12} \mathrm{H}_{26}+12.5 \mathrm{O}_{2}  \;  \rightarrow \;  12 \mathrm{CO}+13 \mathrm{H}_{2} \mathrm{O},
\end{array}
\end{equation}
\begin{equation}
\begin{array}{l}
\mathrm{CO}+0.5 \mathrm{O}_{2} \; \rightleftharpoons \; \mathrm{CO}_{2}.
\end{array}
\end{equation}
The reaction rates (in cgs units) are expressed in Arrhenius form
\begin{equation}
\begin{array}{l}
\mathcal{Q}_1 = \mathcal{A}_1 \exp [-31944 / (\mathcal{R} T) ] \mathcal{C}_{\mathrm{C}_{12} \mathrm{H}_{26}}^{0.25}
\mathcal{C}_{\mathrm{O}_{2} }^{1.25}
\end{array}
\end{equation}
\begin{equation}
\begin{array}{l}
\mathcal{Q}_2 = 3.98 \times 10^{14} \exp [-40000 / (\mathcal{R} T) ]
\mathcal{C}_{\mathrm{C} \mathrm{O} }^{}
\mathcal{C}_{\mathrm{H}_{2}\mathrm{O}  }^{0.5}
\mathcal{C}_{\mathrm{O}_{2} }^{0.25} \ldots \\
-
 5 \times 10^{8} \exp [-40000 / (\mathcal{R} T) ] \mathcal{C}_{\mathrm{C} \mathrm{O}_{2}}^{}
\end{array}
\end{equation}
The logarithm of the Arrhenius pre-exponential factor of the first reaction varies according to the local fresh gas condition dynamically through
\begin{equation}\label{eq:hakim_param}
\begin{array}{l}
\ln \mathcal{A}_1=\theta_{0}+\theta_{1} \exp ( \theta_{2} \phi_0 )+\theta_{3} \tanh [ (\theta_{4}+\theta_{5} \phi_0 ) T_{0}+\theta_{6} ].
\end{array}
\end{equation}
\begin{table}[!htb]
\centering
\caption{The optimal parameters for evaluation of the pre-exponential factor of n-dodecane reaction in the reduced mechanism of Hakim et al. \cite{Hakim2016modeling}.}
\label{table:reaction_params}
\begin{tabular}[]{|c|c|c|c|c|c|c|}
\hline
$\theta_{0}$ & $\theta_{1}$ & $\theta_{2}$ & $\theta_{3}$ & $\theta_{4}$ & $\theta_{5}$ & $\theta_{6}$ \\
\hline
27.38 & -2.13 & -2.05 & 1.89 & -0.01 & 2.87e-4 & 8.43 \\
\hline
\end{tabular}
\end{table}
The {best} possible values of $\theta_{0}$ to $\theta_{6}$ {found} by matching the ignition delay time with the skeletal mechanism of Narayanaswamy et al.~\cite{Narayanaswamy2014} are listed in  Table~\ref{table:reaction_params}.  {The local fresh gas temperature $T_0$ and equivalence ratio $\phi_0$ can be estimated using the local value of mixture fraction~\cite{Hakim2016LES}.}

\begin{figure}[!htb]
\centering
\includegraphics[height=44mm]{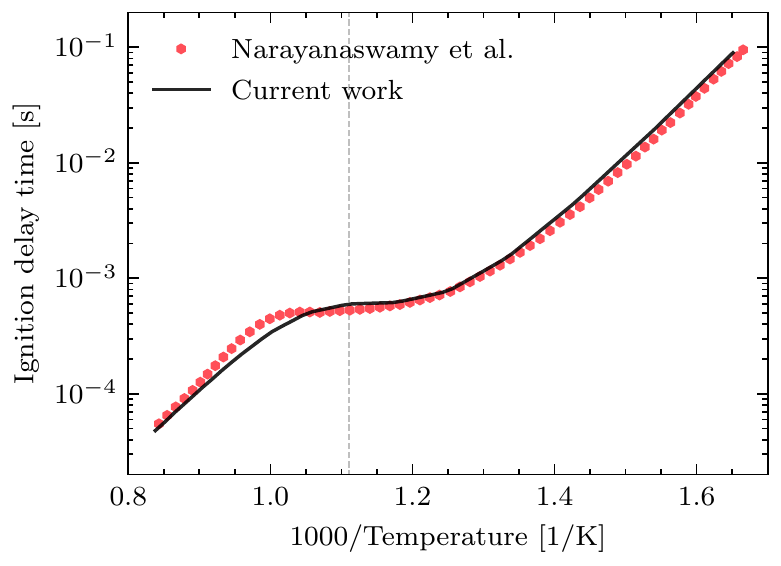}
\caption{Ignition delay time for the skeletal mechanism of Narayanaswamy et al.~\cite{Narayanaswamy2014} and the current work using the reduced mechanism of Hakim et al.~\cite{hakim2018probabilistic} for the stoichiometric mixture of the fuel and oxidizer of ECN Spray-A.}\label{fig:autoignitiontime}
\end{figure}
Fig.~\ref{fig:autoignitiontime} shows an excellent agreement between the ignition delay time predicted by the reduced two-step \num{6}-species reaction mechanism of Hakim et al.~\cite{hakim2018probabilistic}  used in this study, and the skeletal mechanism of Narayanaswamy et al.~\cite{Narayanaswamy2014} with \num{876} reactions and \num{164} species. The ignition delay time is reported as the elapsed time until the initial temperature passes \SI{400}{\kelvin} raise in a {homogeneous constant-volume reactor}. The fluid is a stoichiometric mixture between the fuel and oxidizer with the composition of the reacting Spray-A case.  
The volume of the reactor is set according to the initial temperature, stoichiometric composition, and pressure of \SI{60}{\bar} using RKPR EOS. Table~\ref{table:criticalvals} lists the critical values and the acentric factor of the species required for the simulations of ECN Spray-A with this reduced reaction mechanism.
\begin{table}[!htb]
\centering
\caption{Critical properties and acentric factor of the species.}\label{table:criticalvals}
\begin{tabular}{| c |c |c| c| c| }
\hline
Species & $T_c\;\textnormal{[K]}$ & $p_c\;\textnormal{[bar]}$ & $Z_c\;\textnormal{[-]}$ & $\omega\;\textnormal{[-]}$ \\
\hline
$\mathrm{C}_{12} \mathrm{H}_{26}$ & 658.0 & 18.20 & 0.251 & 0.576 \\
$\mathrm{O}_{2}$ & 154.6 & 50.43 & 0.288 & 0.022 \\
$\mathrm{N}_{2}$ & 126.2 & 34.00 & 0.289 & 0.038 \\
$\mathrm{C} \mathrm{O}_{2}$ & 304.2 & 73.83 & 0.274 & 0.224 \\
$\mathrm{H}_{2} \mathrm{O}$ & 647.1 & 220.6 &  0.229 & 0.345 \\
$\mathrm{C} \mathrm{O}$ & 132.9 & 34.99 & 0.299 & 0.048 \\
\hline
\end{tabular}
\end{table}

\section{Numerical Results}

In this section, we use experimental reference data of the reacting and non-reacting ECN Spray-A in order to evaluate and validate the proposed multiphase thermodynamics and real-fluid finite rate chemistry model for transcritical fuel sprays.

\subsection{Non-reacting case}
\begin{figure}[!htb]
\centering
\includegraphics[width=77mm]{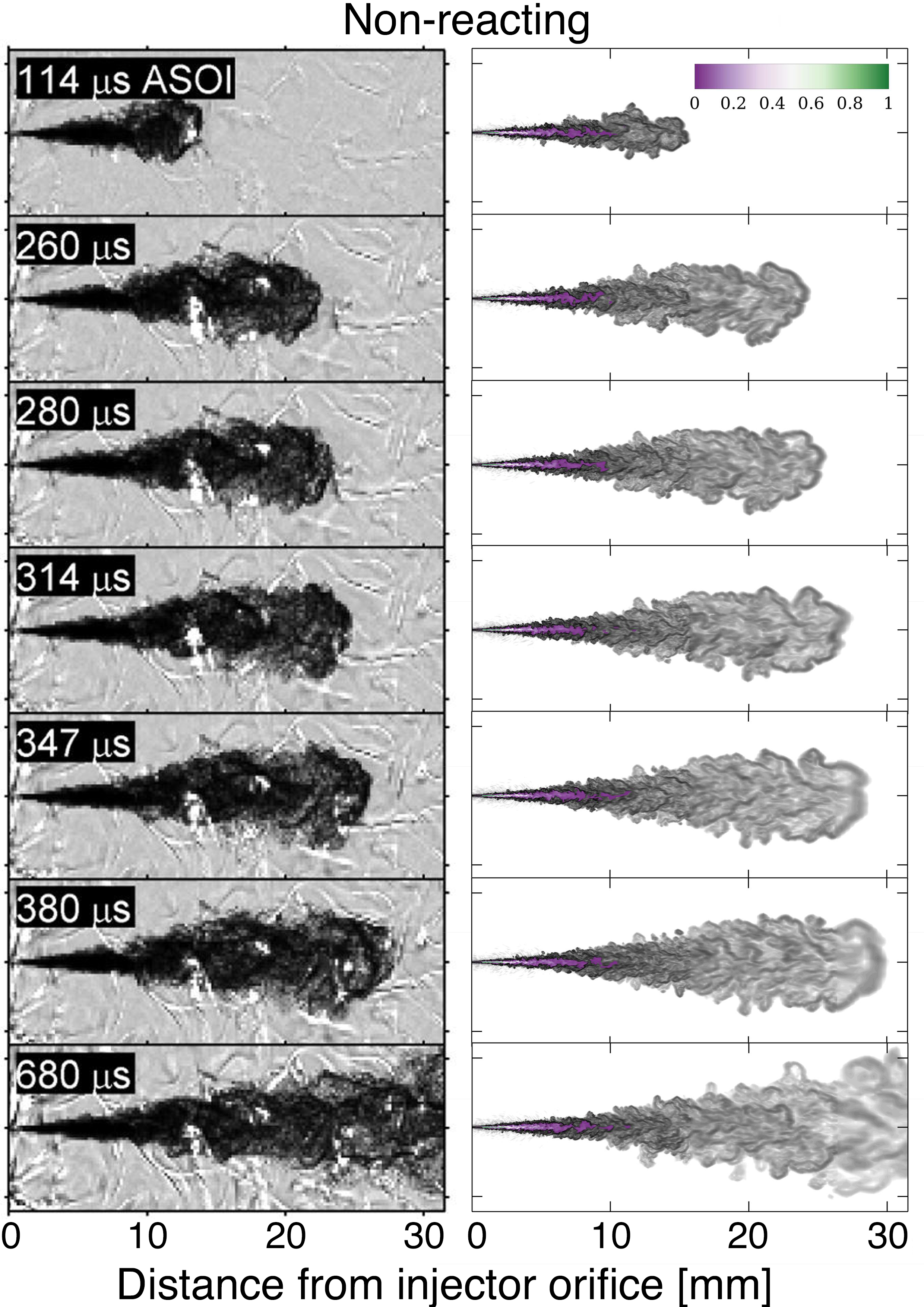}
\caption{Time sequence snapshots of non-reacting Spray-A using Schlieren images~\cite{skeen2015simultaneous} and predictions by the LES-MT method. The background contour of LES data shows the gradient of density, which is overprinted in the two-phase region by liquid volume fraction contours.}\label{fig:inert_snapshots}
\end{figure}

\begin{figure}[!htb]
         \centering
     \begin{subfigure}[c]{60mm}
         \includegraphics[height=44mm]{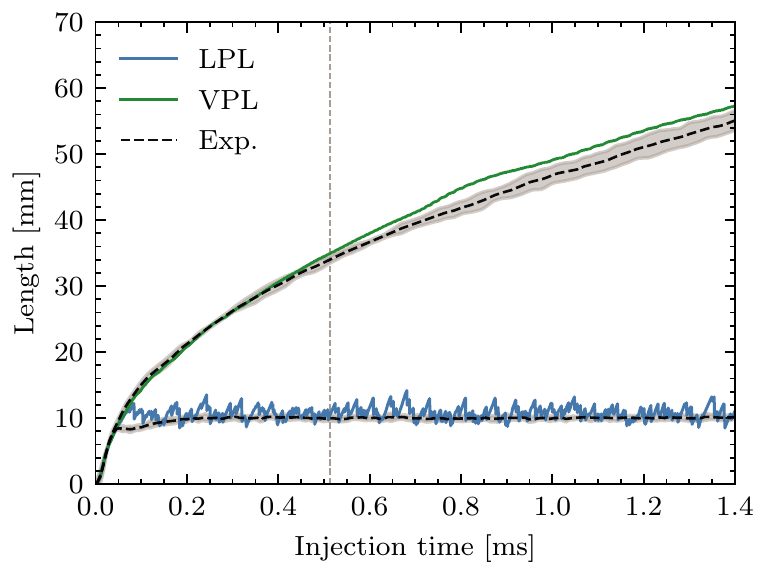}
     \end{subfigure}
     \begin{subfigure}[c]{37mm}
         \includegraphics[width=37mm]{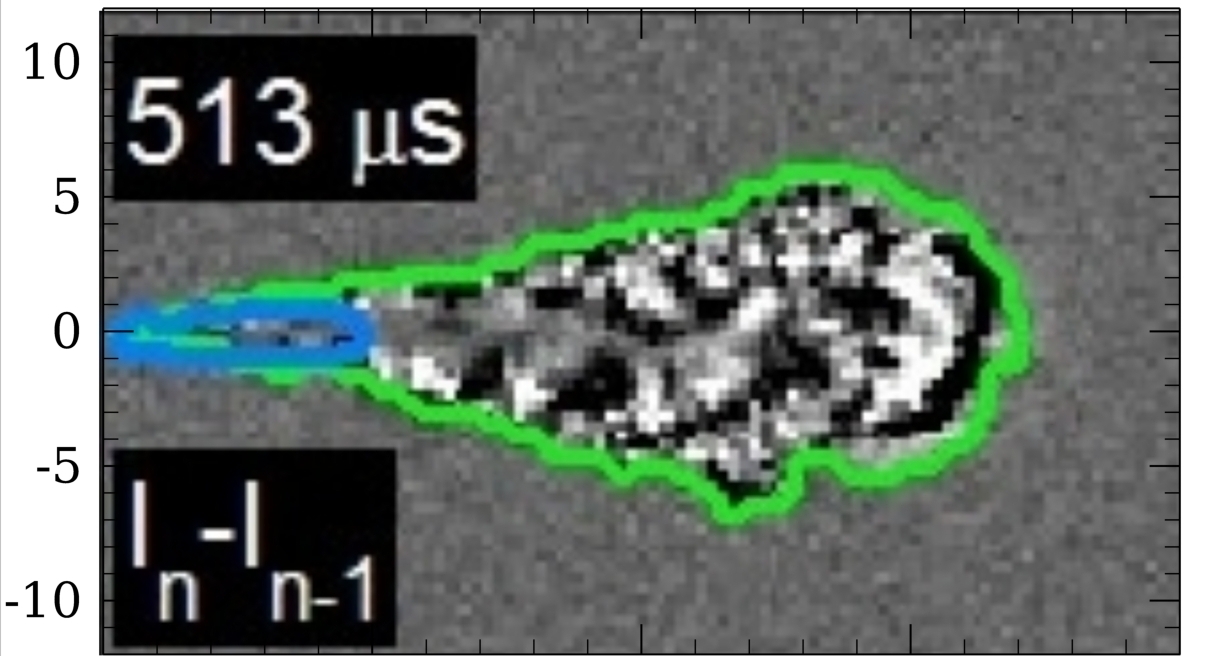}
         \includegraphics[width=37mm]{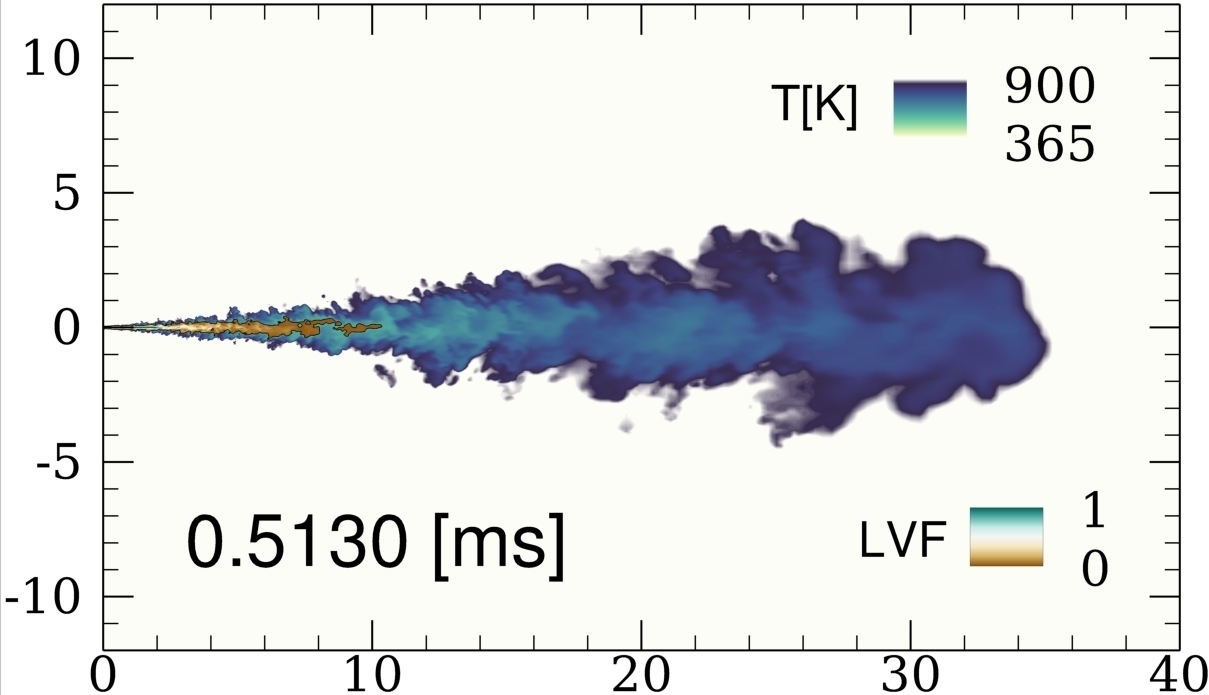}
     \end{subfigure}
\caption{Left is vapor and liquid penetration trajectories for LES-MT and experiment~\cite{Pickett1,Pickett2} for the non-reacting ECN Spray-A. On the right, are an experimental Schlieren image with highlighted vapor and liquid boundaries along with an LES-MT  temperature contour overprinted with LVF in the liquid and two-phase region at the same instance.}\label{fig:inert_penetration}
\end{figure}

In Fig.~\ref{fig:inert_snapshots}, snapshots of inert Spray-A at 7 specific times after the start of injection (ASOI) predicted by the LES-MT model (right column) are compared with experimental Schlieren images~\cite{skeen2015simultaneous}. The non-reacting jet first penetrates in axial direction and spreads in radial direction while evaporating and mixing with the hot chamber gas, and then  maintains an approximately constant radius after some downstream distance from the injector nozzle.
In the experimental Schlieren images, the saturated dark regions represent the liquid phase. For LES-MT snapshots, we show colored contours of the liquid volume fraction (LVF), which also indicate the amount of the liquid phase, and density gradients in the single-phase gaseous regions, where LVF=0. The liquid penetration length is reported to be about \SI{10}{\milli\meter} in the experiment, which is in excellent agreement with our numerical results, see also Fig.~\ref{fig:inert_penetration}.  

Figure~\ref{fig:inert_penetration} shows the temporal evolution of the liquid penetration length (LPL) and the vapor penetration length (VPL) for the LES-MT numerical simulation and experimental measurements~\cite{Pickett1,Pickett2}. For the LES-MT, LPL and VPL are defined as the maximum axial locations with a LVF of \SI{15}{\percent} and a mixture fraction of \SI{0.01}{\percent}, respectively, similar as in previous studies~\cite{matheis2016multi}. The LPL signals of LES and experiment are in excellent agreement. For the VPL, we observe excellent match up to about \SI{0.6}{\milli\second}. At later times, the simulation predicts slightly larger values that measured in the experiment. This could be do to the coarsened mesh at those locations far from the nozzle, or due to measurement uncertainties. The estimated uncertainty of the experimental measurements is indicated by the gray-shaded area and significantly increases with time. On the right side of the same figure, we also compare an experimental snapshot with highlighted liquid and vapor boundaries with a numerical visualization. The agreement is good, considering that both snapshots are instantaneous samples of independent realizations of highly turbulent flows.

\begin{figure}[!htb]
\centering
\includegraphics[height=44mm]{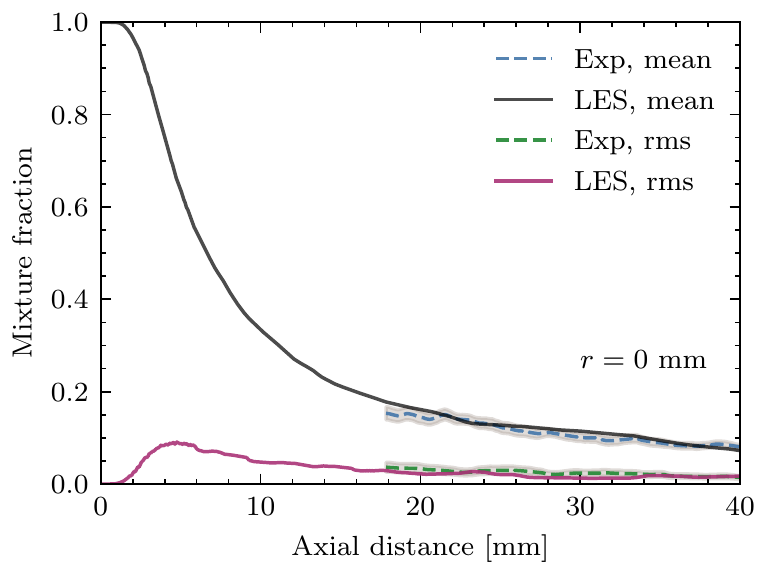}
\includegraphics[height=44mm]{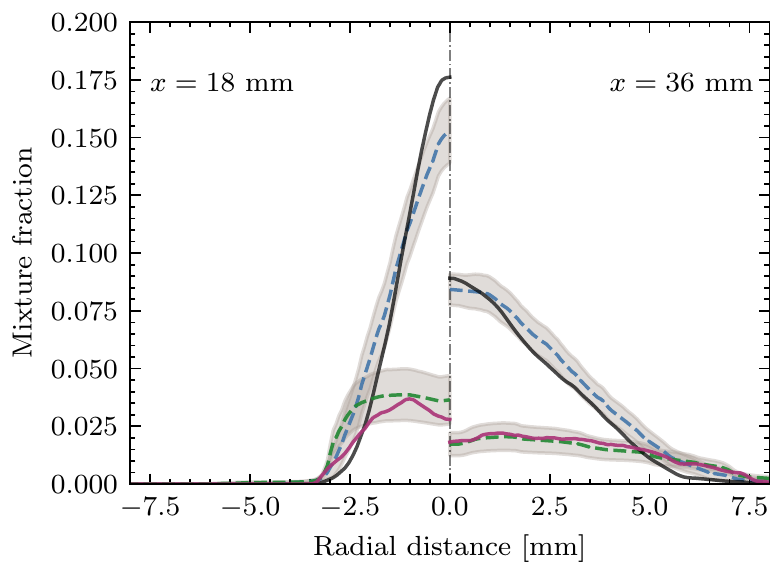}
\caption{Axial profile and radial profiles of mean and RMS fluctuations of the mixture fraction computed using LES-MT method in comparison with Rayleigh scattering measurements~\cite{Pickett3} for the non-reacting Spray-A.}\label{fig:inert_statistics}
\end{figure}

We present ensemble averaged profiles of the mixture fraction on the center line and at two downstream locations in Fig.~\ref{fig:inert_statistics}. The statistics have been computed by ensemble averaging LES data collected at \num{8} circumferential sections every \SI{0.5}{\micro\second} during the time interval highlighted in Fig.~\ref{fig:transientvelocity}. LES-MT results follow the measured axial profiles very well. At the first station, $x=$\SI{18}{\milli\meter}, the n-dodecane mass fraction on the jet axis is overestimated by the LES compared to the experiment; however, the LES data fully agrees with the experimental data further downstream, which can be also seen from the radial profiles at $x=$\SI{36}{\milli\meter}.

\begin{figure}[!htb]
         \centering
     \begin{subfigure}[b]{0.31\linewidth}
         \centering
         \includegraphics[width=\linewidth]{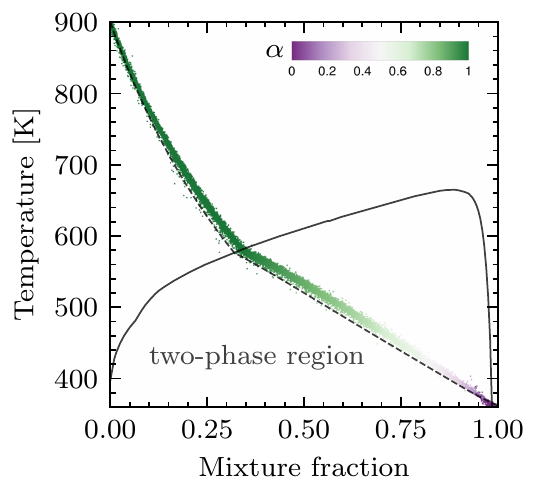}
     \end{subfigure}
     \begin{subfigure}[b]{0.31\linewidth}
         \centering
         \includegraphics[width=\linewidth]{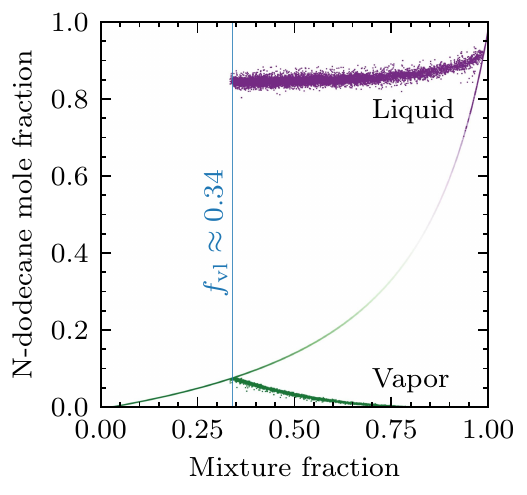}
     \end{subfigure}
     \begin{subfigure}[b]{0.31\linewidth}
         \centering
         \includegraphics[width=\linewidth]{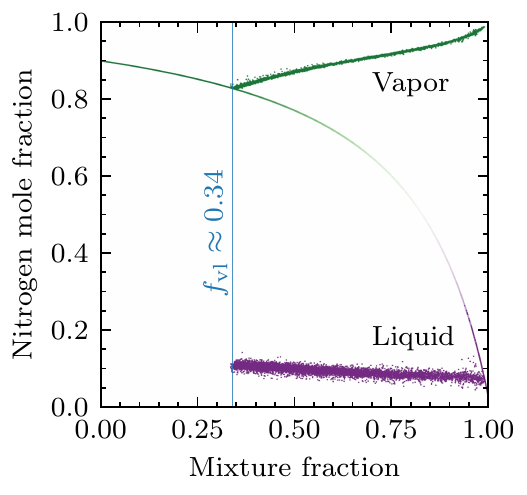}
     \end{subfigure}
\caption{Scattered data depicts the LES results of the non-reacting Spray-A at \SI{680}{\micro\second} and they are colored by vapor mole fraction values. Temperature-mixture fraction diagram for the quaternary mixture is shown with  temperature variations during the isenthalpic process (dashed line) and two phase region boundaries at \SI{60}{\bar}, N-dodecane and Nitrogen mole fractions  diagrams are shown for the overall as well as the saturated liquid and vapor phases.}\label{fig:inert_mixturefractionspace}
\end{figure}

Figure~\ref{fig:inert_mixturefractionspace} shows global scatter plots of the temperature and the mole fractions of the major species n-dodecane and nitrogen as a function of the mixture fraction. Each point represents an instantaneous local state of the resolved LES-MT flow field for the non-reacting case at \SI{680}{\micro\second}. The data points are colored with the molar vapor fraction in a way that green is completely vapor and purple is completely liquid. The two-phase region  at the nominal operating pressure of \SI{60}{\bar} is indicated in the temperature/mixture-fraction diagram. A hypothetical temperature profile predicted by multiphase thermodynamics assuming isenthalpic mixing at \SI{60}{\bar} is also shown. Two-phase boundary and isenthalpic curve cross each other at the VLE mixture fraction of about \num{0.34} for this case using RKPR EOS. The LES data points follow the isenthalpic line strikingly well. This is explained in more detail in Refs.~\cite{matheis2016multi,Matheis2018}. In the two phase region, liquid and vapor phases have different molar composition, as determined by phase-equilibrium computations. The vapor and liquid compositions are shown for the two major species, n-dodecane and nitrogen, in the mole-fraction/mixture-fraction diagrams in Fig.~\ref{fig:inert_mixturefractionspace}. The diagrams indicate that the saturated liquid contains only about \SI{90}{\percent} n-dodecane and the rest is mostly nitrogen. This is a consequence of high solubility at transcritical pressures, and questions the standard pure-fuel assumption made for liquid droplet in traditional LPT methods.

\begin{figure}[!htb]
         \centering
     \begin{subfigure}[b]{0.4\linewidth}
         \centering
         \includegraphics[width=\linewidth]{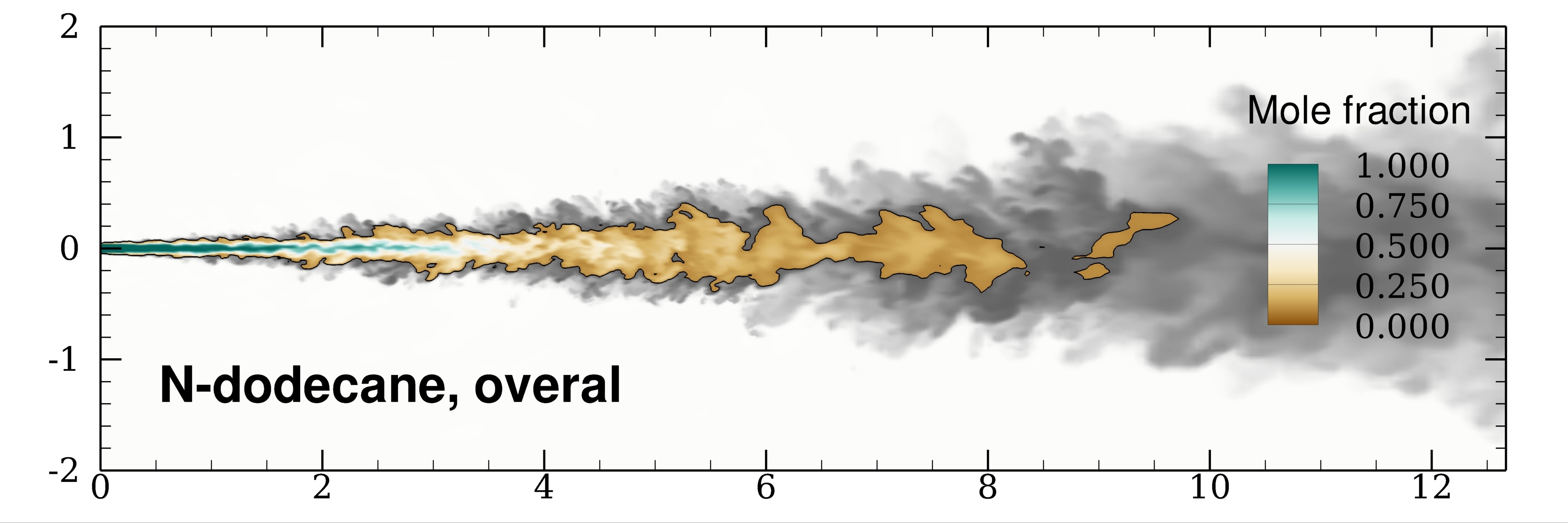}
         \includegraphics[width=\linewidth]{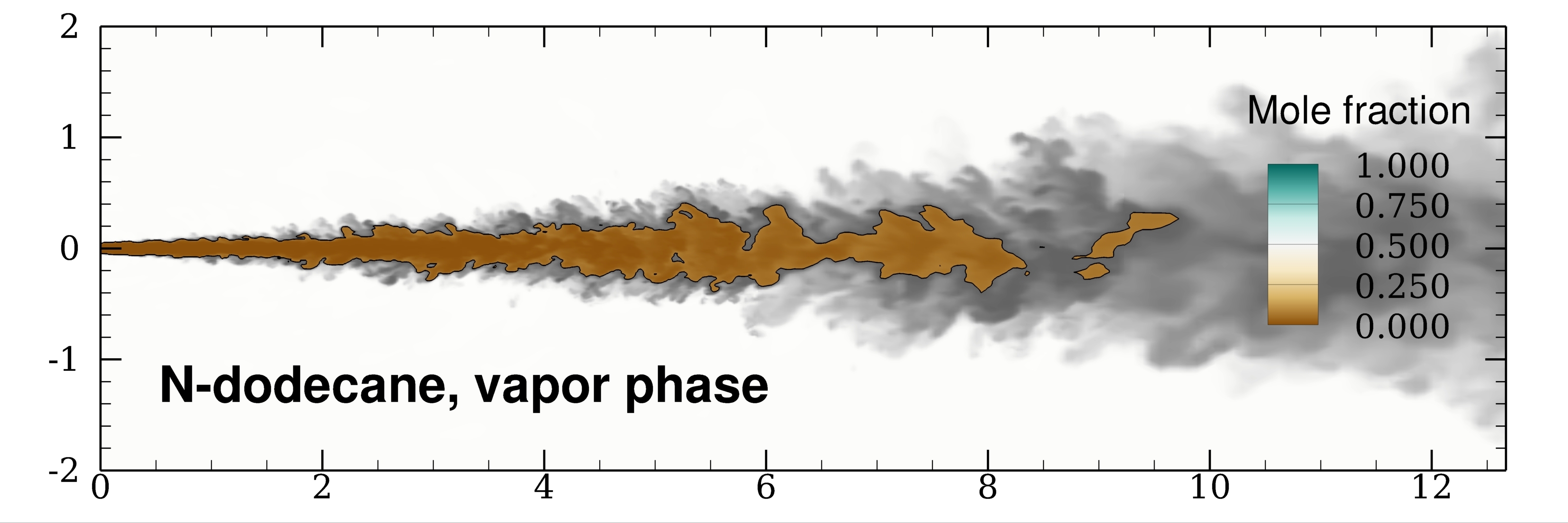}
         \includegraphics[width=\linewidth]{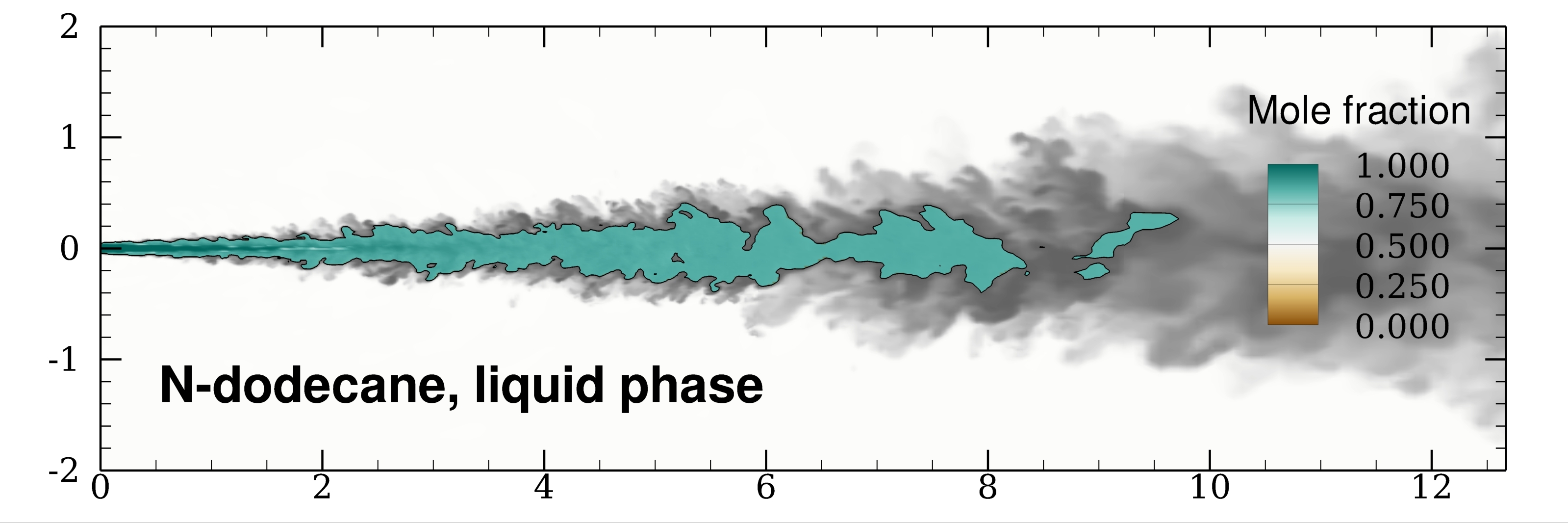}
     \end{subfigure}
     \begin{subfigure}[b]{0.4\linewidth}
         \centering
         \includegraphics[width=\linewidth]{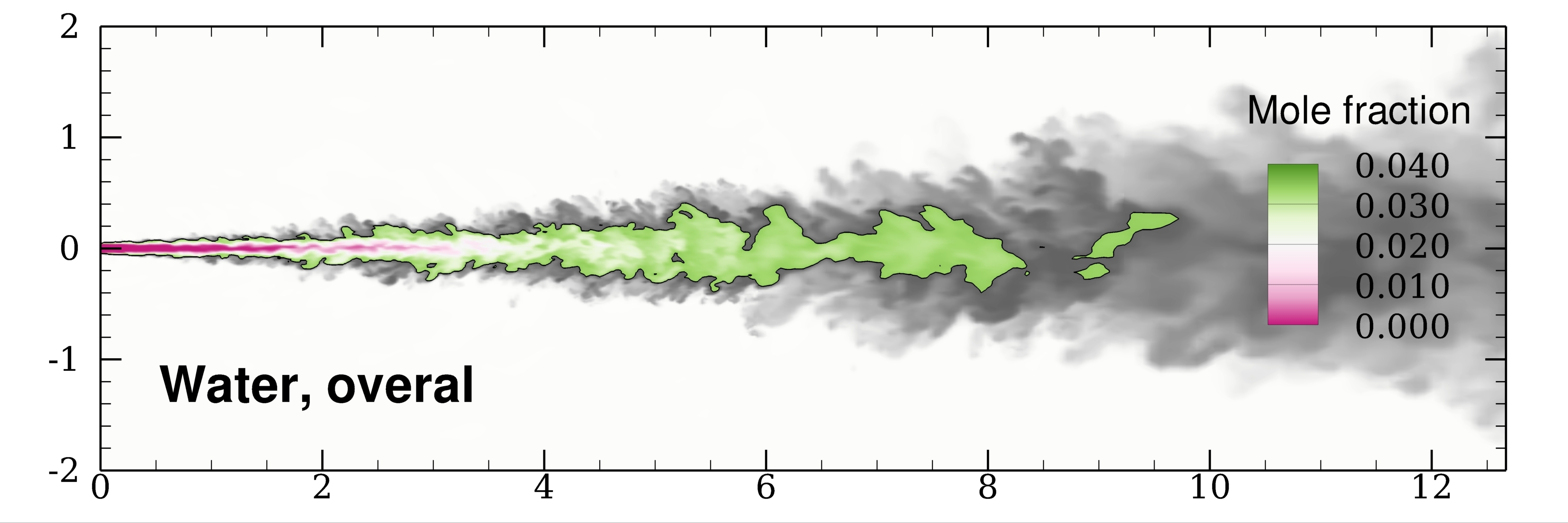}
         \includegraphics[width=\linewidth]{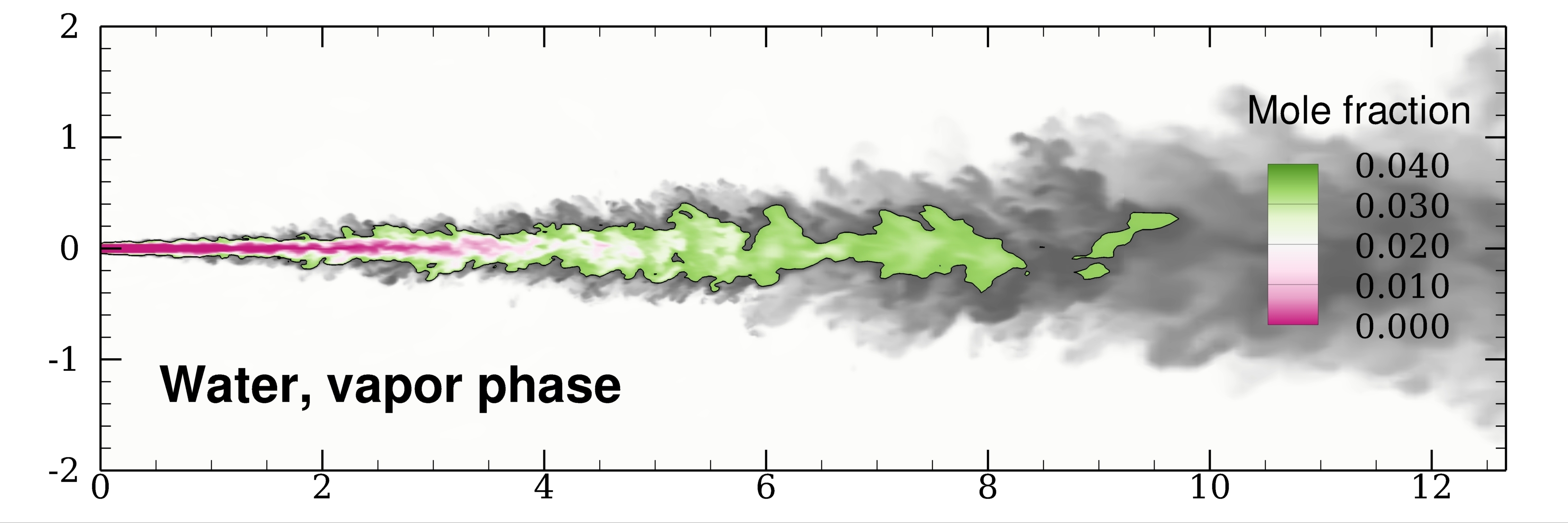}
         \includegraphics[width=\linewidth]{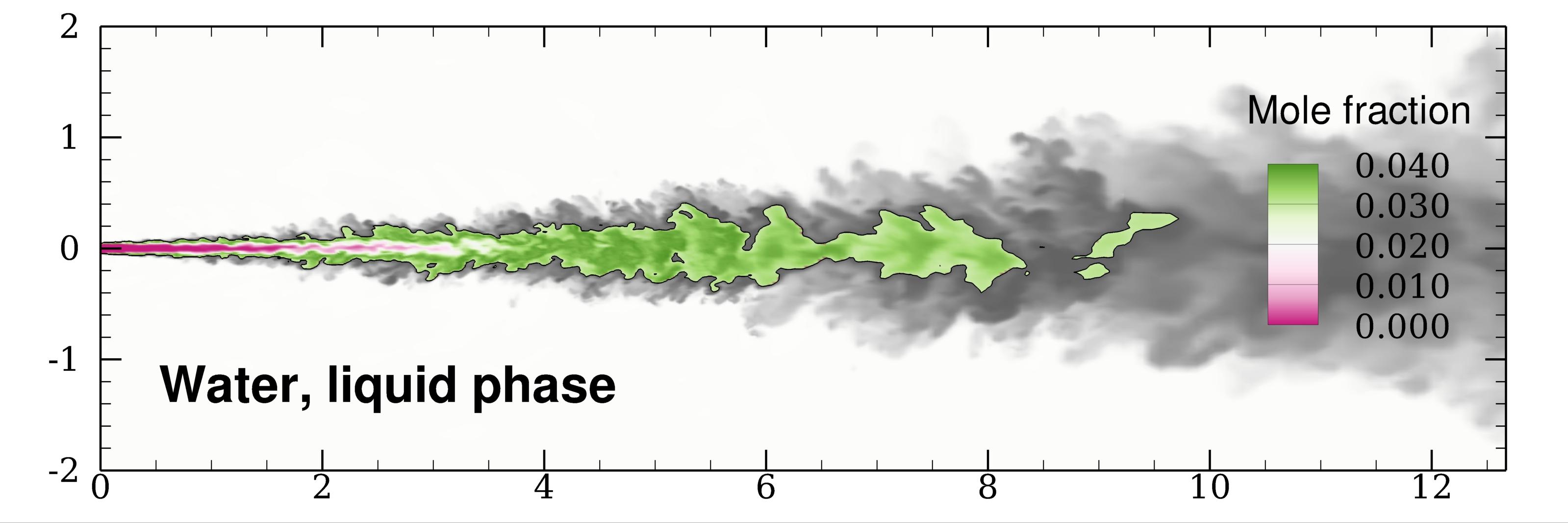}
     \end{subfigure}
     \vskip0.5cm
     \begin{subfigure}[b]{0.4\linewidth}
         \centering
         \includegraphics[width=\linewidth]{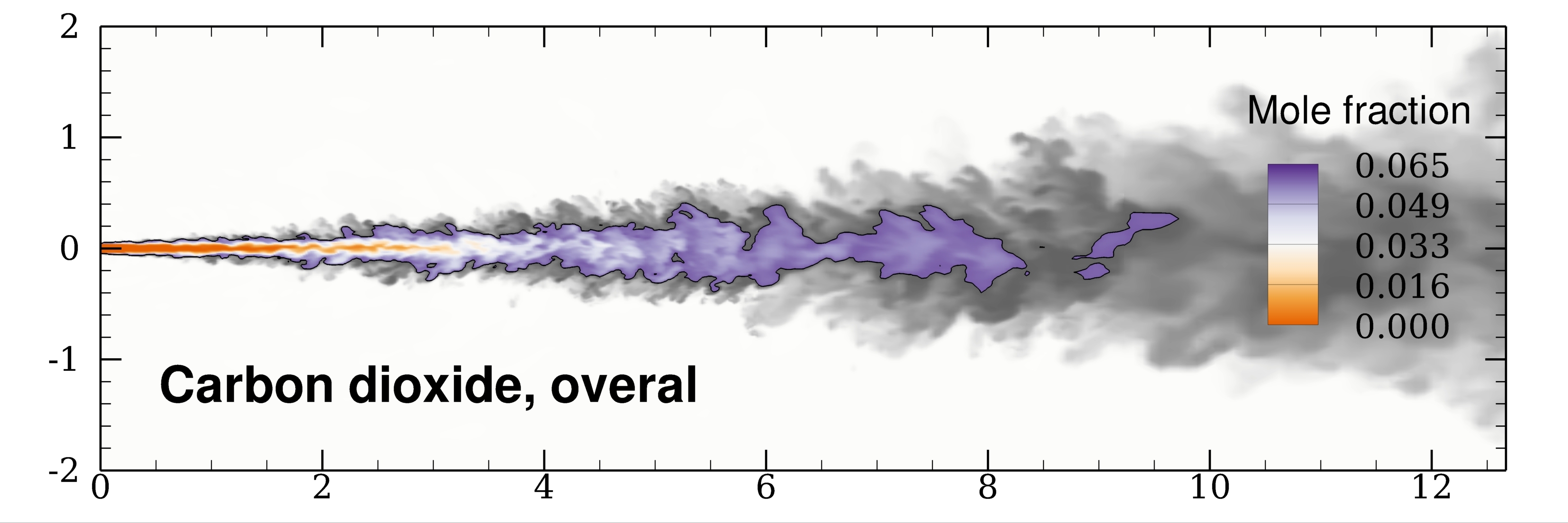}
         \includegraphics[width=\linewidth]{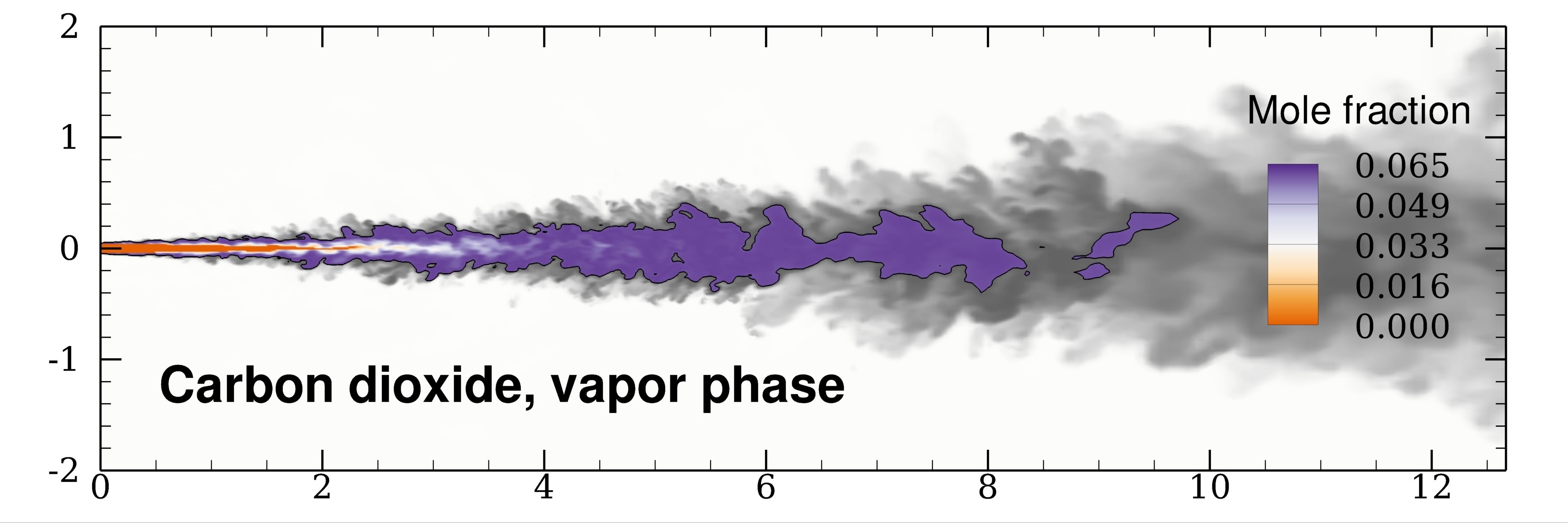}
         \includegraphics[width=\linewidth]{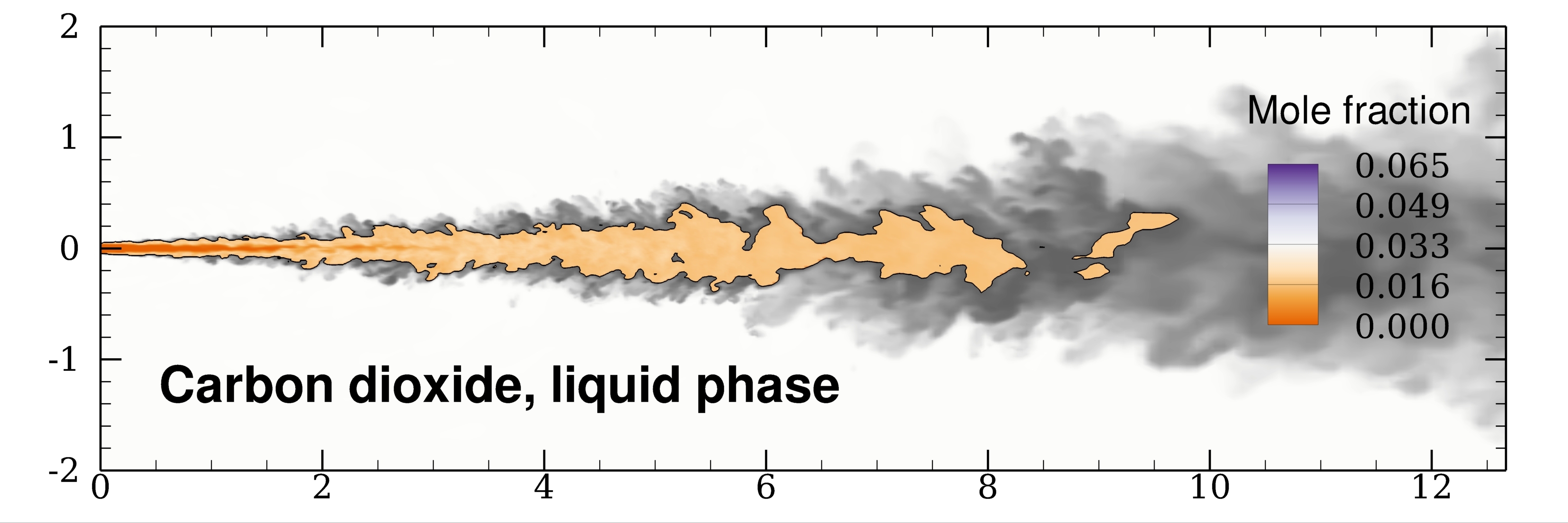}
     \end{subfigure}
     \begin{subfigure}[b]{0.4\linewidth}
         \centering
         \includegraphics[width=\linewidth]{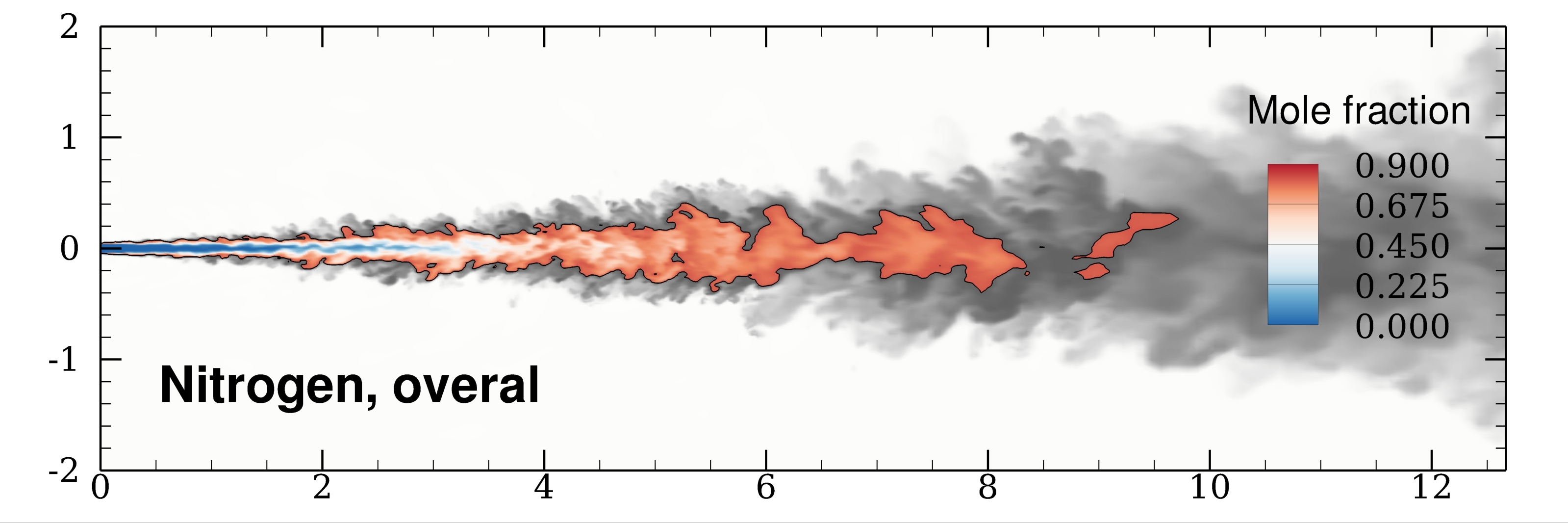}
         \includegraphics[width=\linewidth]{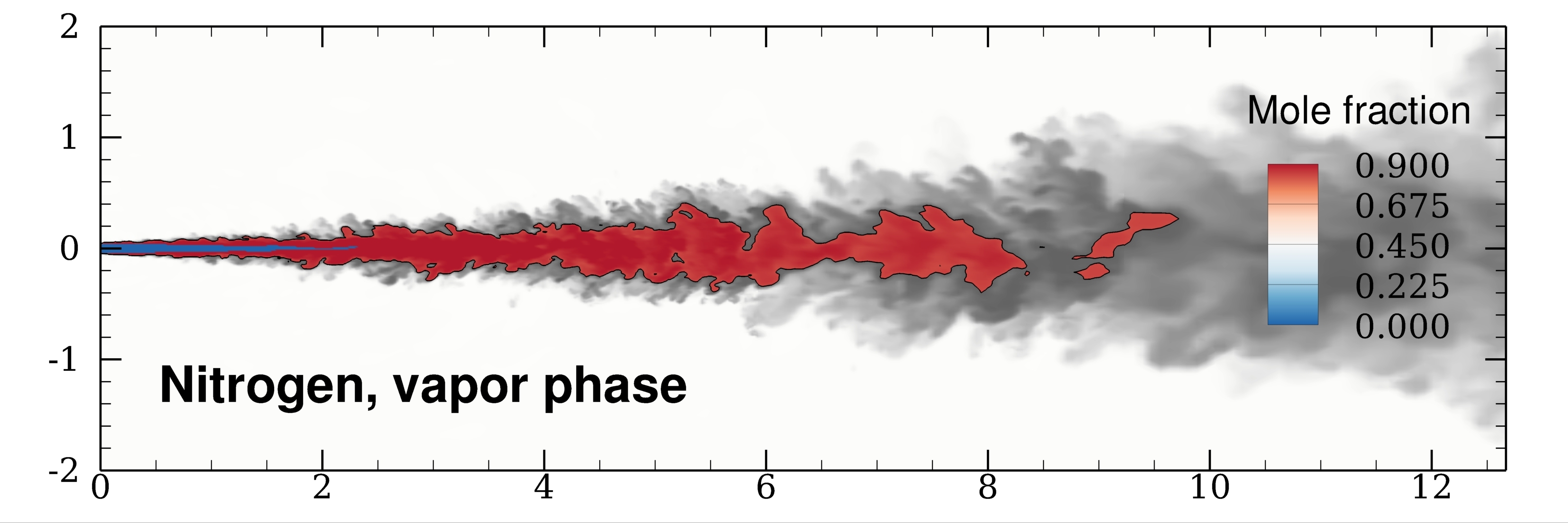}
         \includegraphics[width=\linewidth]{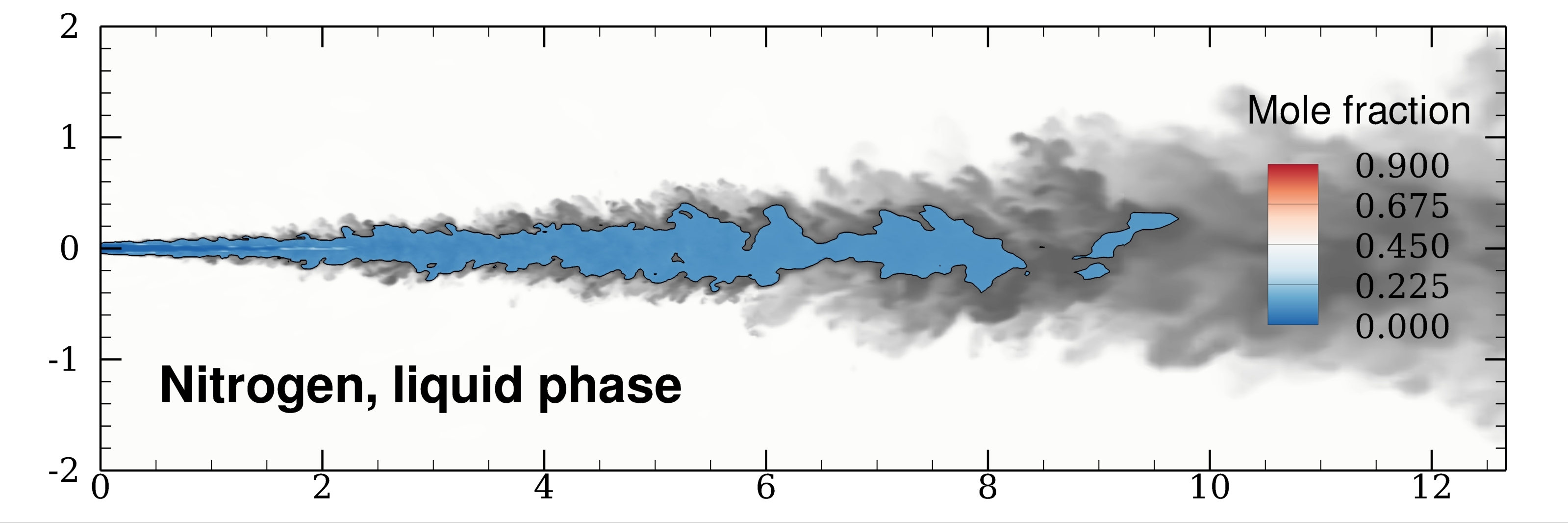}
     \end{subfigure}
     \caption{Instantaneous snapshots for the vapor-liquid molar composition for the non-reacting case. The background contour represents the temperature field from dark to light shades. In the two-phase region, contours of the species mole fraction of saturated liquid, saturated vapor and overall mixture are presented.}\label{fig:inert_composition}
\end{figure}

Figure~\ref{fig:inert_composition} shows contours of the molar composition of the overall mixture as well as the compositions of the liquid and vapor phases for all species. The figure indicates the spatial evolution of the partial mixing of fuel and environment controlled by the multiphase thermodynamics. While the heavy molecules of n-dodecane are about \SI{90}{\percent} of the total moles of the liquid core, they are in minority in the vapor phase, which consists mainly of light nitrogen molecules. The fuel rich liquid core extends from the injector nozzle up to about \SI{3}{\milli\meter}. Far from the nozzle and close to the tip of the liquid core, where the vapor mole fraction is close to unity, the mixture composition is close to the saturated vapor phase and, accordingly, the mixture fraction is close to \SI{0.34}. In contrast to nitrogen and n-dodecane, carbon dioxide and water have roughly the same composition in the liquid and vapor phases.

\subsection{Reacting case}
\begin{figure}[!htb]
\centering
\includegraphics[width=77mm]{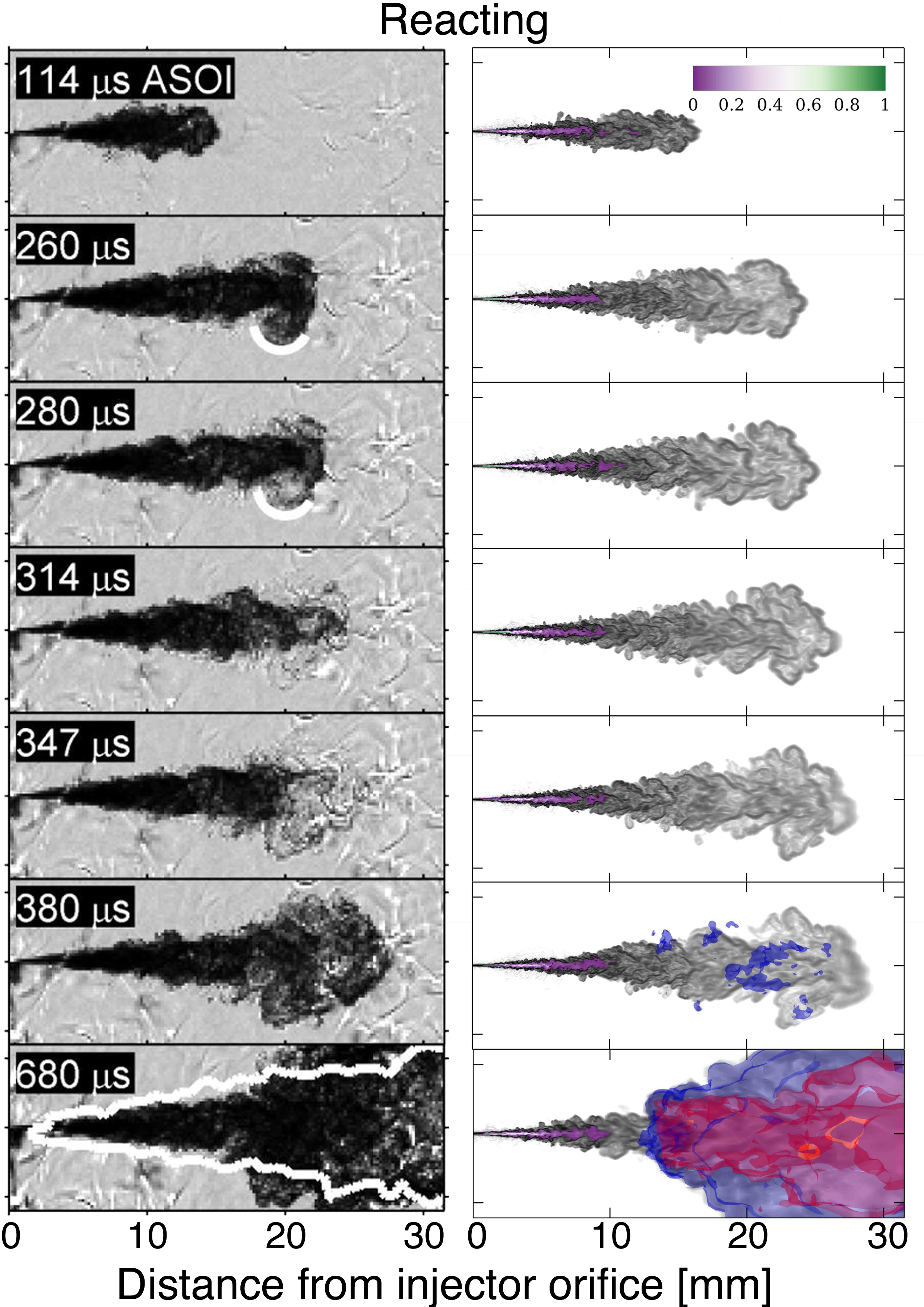}
\caption{Time sequence snapshots of reacting Spray-A for experiment (Schlieren images on the right~\cite{skeen2015simultaneous}) and LES-MT (left). For the LES-MT, background contours of the density gradient are overprinted by liquid volume fraction contours in the two-phase region, and blue and red iso-surfaces for temperatures of \SI{920}{\kelvin} and \SI{2000}{\kelvin} enclose the regions of low and high temperature reactions.}\label{fig:reacting_snapshots}
\end{figure}

A fist impression of the time evolution of the reacting ECN Spray-A case is provided in Fig.~\ref{fig:reacting_snapshots} based on experimental Schlieren images ~\cite{skeen2015simultaneous} on the left side and corresponding snapshots of LES-MT solution on the right side. Similar to the inert test case, the liquid phase appears as a saturated black region in the experimental Schlieren images and contours of the LVF show the predicted amount of liquid for the LES-MT. There is a very good agreement between the experiment and simulation as before. The jet develops very similar to non-reacting spray up to the auto-ignition.  The ignition delay time is approximately \SI{400}{\micro\second} for both experiment and simulation. Around to the ignition time,  low-temperature reactions are activated in a significant portion of the vaporized fuel. This is detectable between \SI{20} and \SI{25}{\milli\meter} after \SI{314}{\micro\second} via brighter regions in the Schlieren images, and visualized by blue iso-temperature surfaces at \SI{920}{\kelvin} for the LES-MT results.  The high-temperature ignition after the low-temperature reactions created enough radicals and intermediate species. The highly reacting region is highlighted by the red iso-temperature surface at \SI{2000}{\kelvin} for the LES results in the last snapshot at \SI{680}{\micro\second}. The high-temperature ignition is a rapid volumetric process. The abrupt radial expansion of the reacting jet, compared to the inert case, indicates the location of quasi-steady the lift-off length (LOL), which is about \SI{17}{\milli\meter}. We observe good agreement of the experimental LOL with our LES-MT result.

\begin{figure}[!htb]
         \centering
     \begin{subfigure}[c]{60mm}
         \includegraphics[height=44mm]{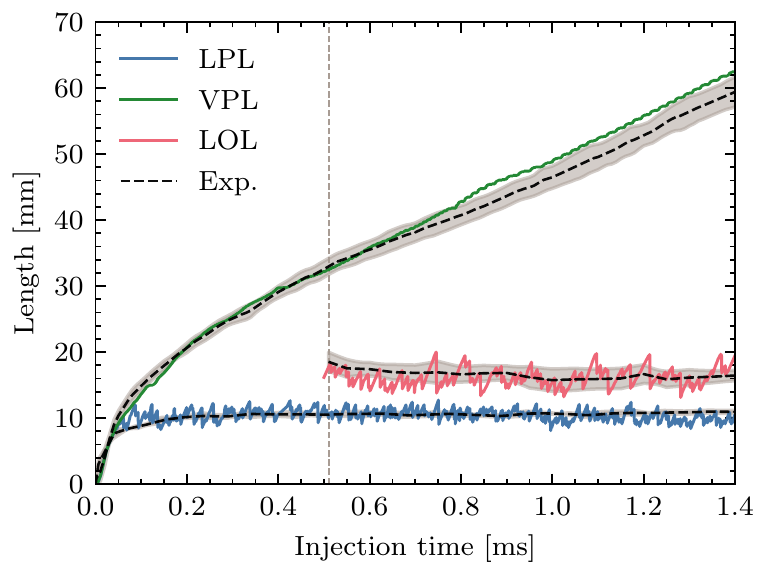}
         %\caption{$y=x$}
         %\label{fig:y equals x}
     \end{subfigure}
     \begin{subfigure}[c]{37mm}
         \includegraphics[width=37mm]{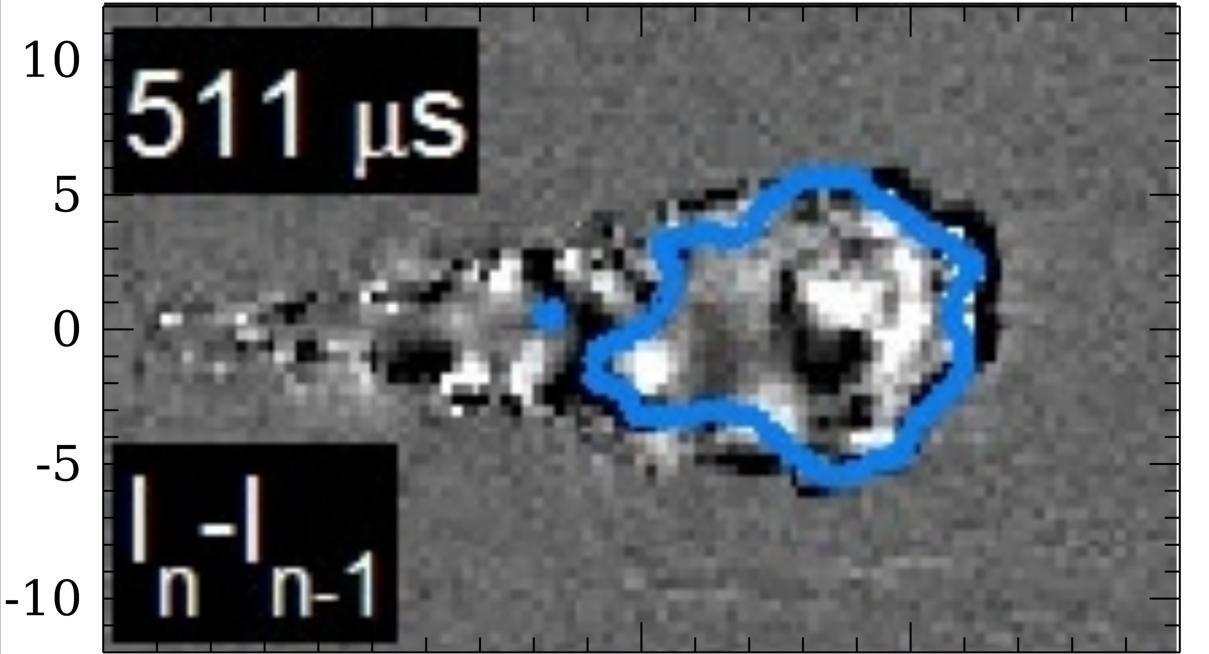}
         \includegraphics[width=37mm]{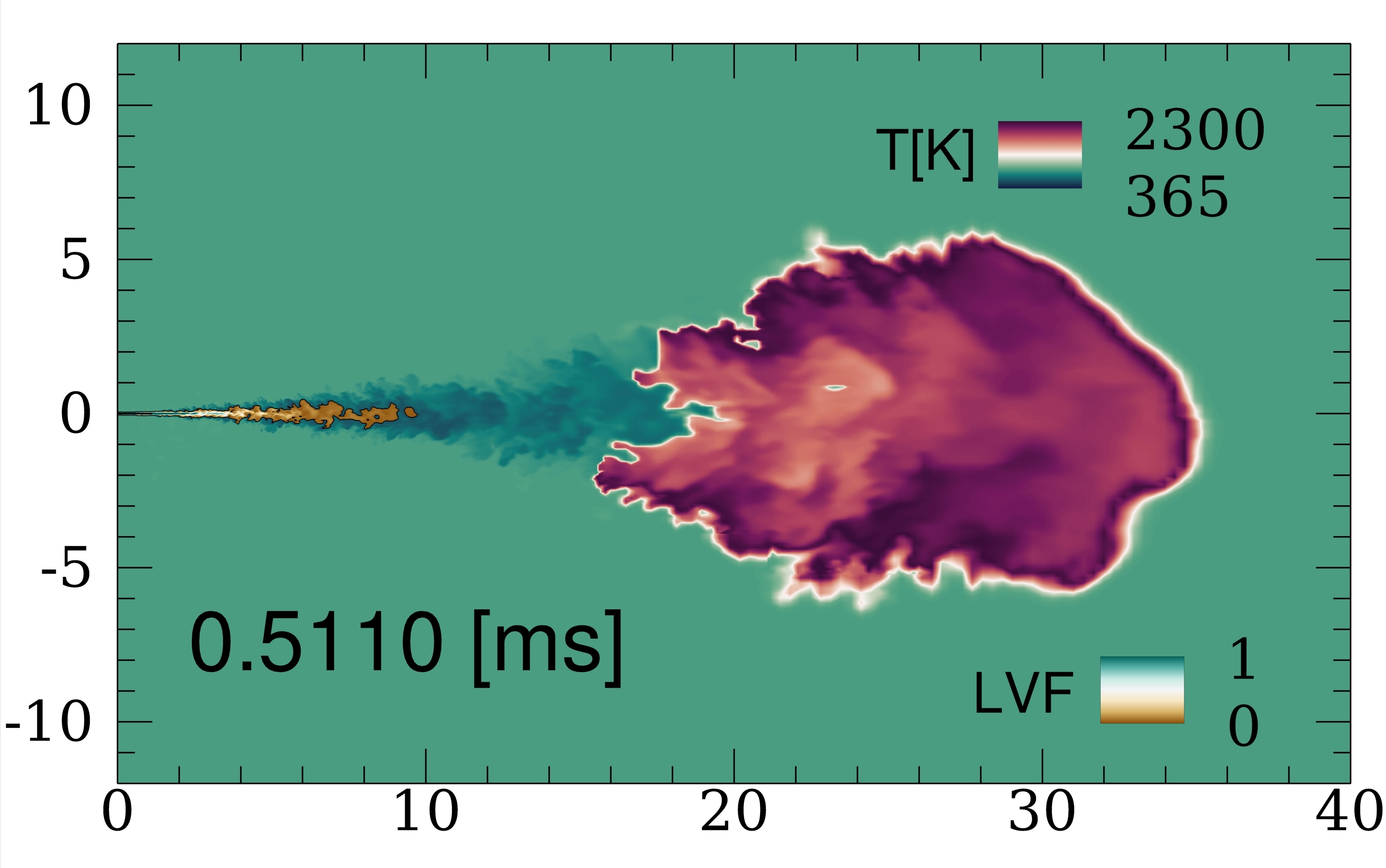}
         %\caption{$y=3sinx$}
         %\label{fig:three sin x}
     \end{subfigure}
\caption{Left is vapor and liquid penetration trajectories predicted by the LES-MT method and measured experimentally ~\cite{Pickett1,Pickett2} for the reacting Spray-A. The two right panels show a sample experimental Schlieren image with highlighted flame boundaries along with an LES-MT temperature contour overprinted with the LVF at the same instant.}\label{fig:reacting_penetration}
\end{figure}

The reacting transcritical jet can be characterised by three important lengths scales: LPL, VPL, and LOL. Figure~\ref{fig:reacting_penetration} shows the time evolution of these lengths for our LES-MT results and the experimental measurements~\cite{Pickett1,Pickett2}. For the LES-MT, LPL and VPL are computed by the method explained above for the non-reacting case. The LOL is computed as the minimum axial location where the temperature exceeds \SI{1800}{\kelvin}. The uncertainty of the experimental measurements is indicated by the gray area around the dotted lines in Fig.~\ref{fig:reacting_penetration}.
The numerically predicted and experimentally measured LPL evolution is in excellent agreement. The VPL evolution is in very good agreement up to about \SI{40}{\milli\meter} and afterwards the values predicted by the simulation are slightly higher than the experimental data. This is consistent with the results for the non-reacting case. The evolution of the flame LOL is again in excellent agreement with the experiment, even though we use a highly reduced reaction mechanism for the simulation.
In order to visualize the flame shape and volume, an experimental snapshot with highlighted high temperature boundaries is compared with the temperature contour plot for an LES-MT snapshot at the the same time instant on the right side of Fig.~\ref{fig:reacting_penetration}. The numerical and experimental flame snapshots are strikingly similar. 

\begin{figure}[!htb]
         \centering
     \begin{subfigure}[b]{0.25\linewidth}
         \centering
         \includegraphics[width=\linewidth]{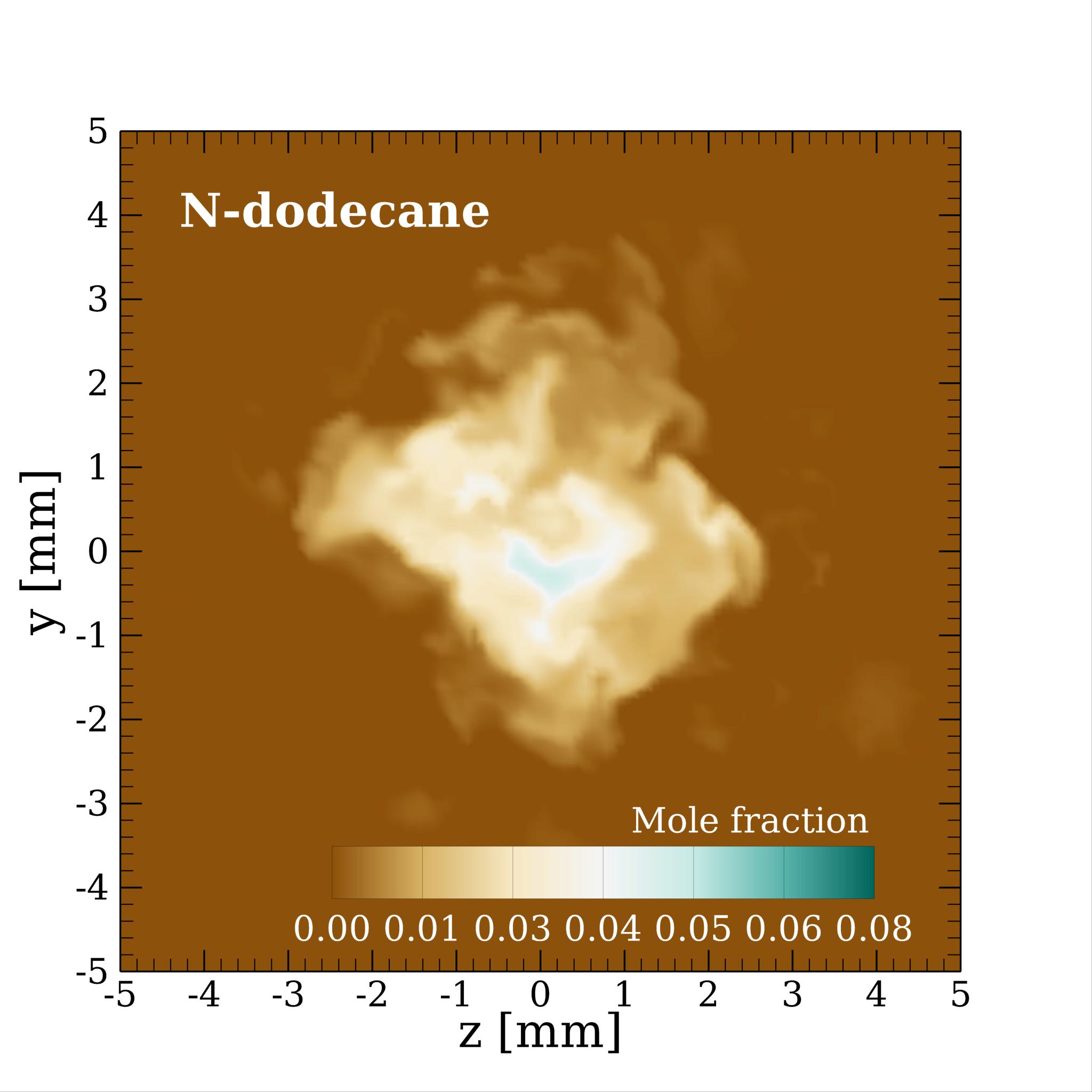}
         %\caption{$y=x$}
         %\label{fig:y equals x}
     \end{subfigure}
     \begin{subfigure}[b]{0.25\linewidth}
         \centering
         \includegraphics[width=\linewidth]{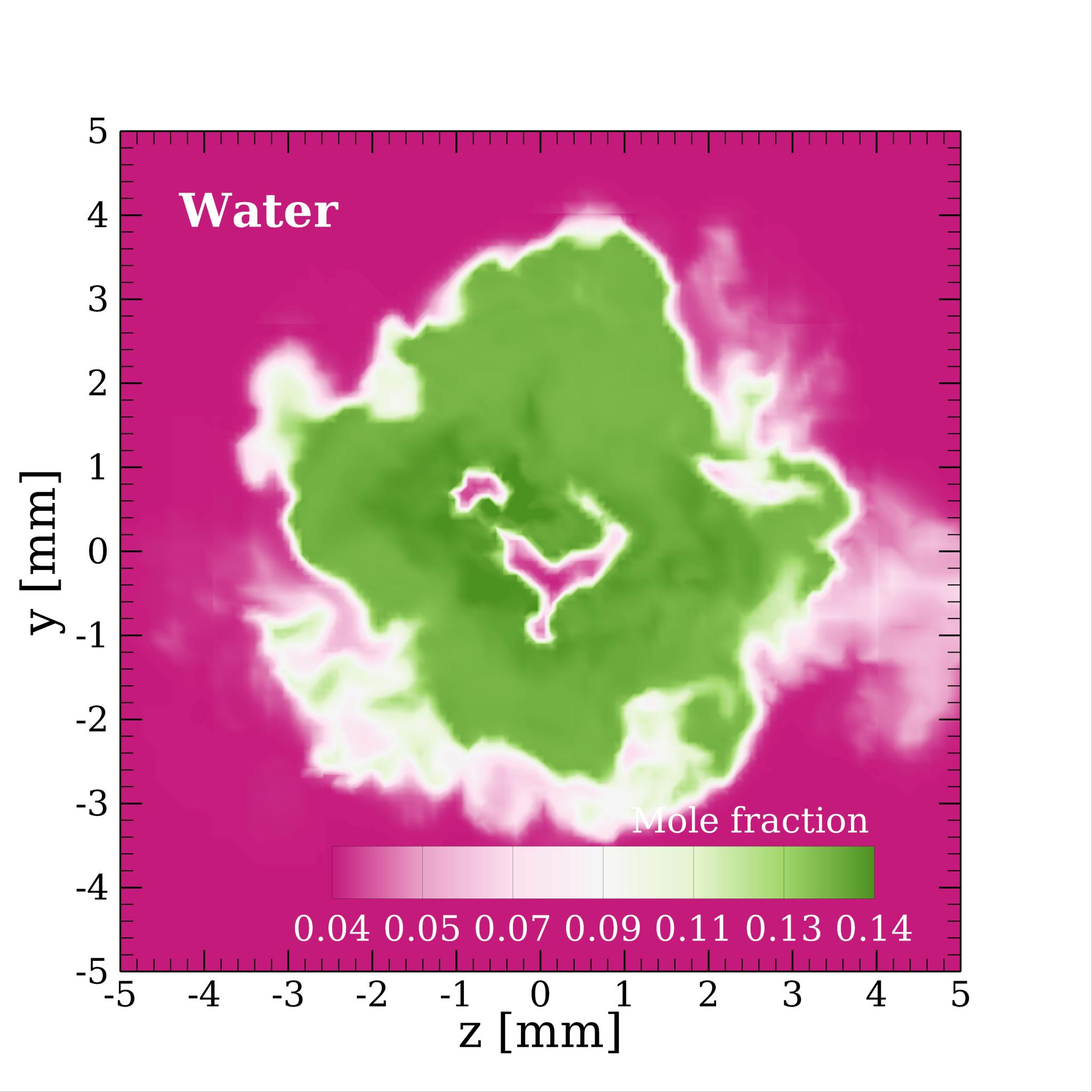}
         %\caption{$y=3sinx$}
         %\label{fig:three sin x}
     \end{subfigure}
     \begin{subfigure}[b]{0.25\linewidth}
         \centering
         \includegraphics[width=\linewidth]{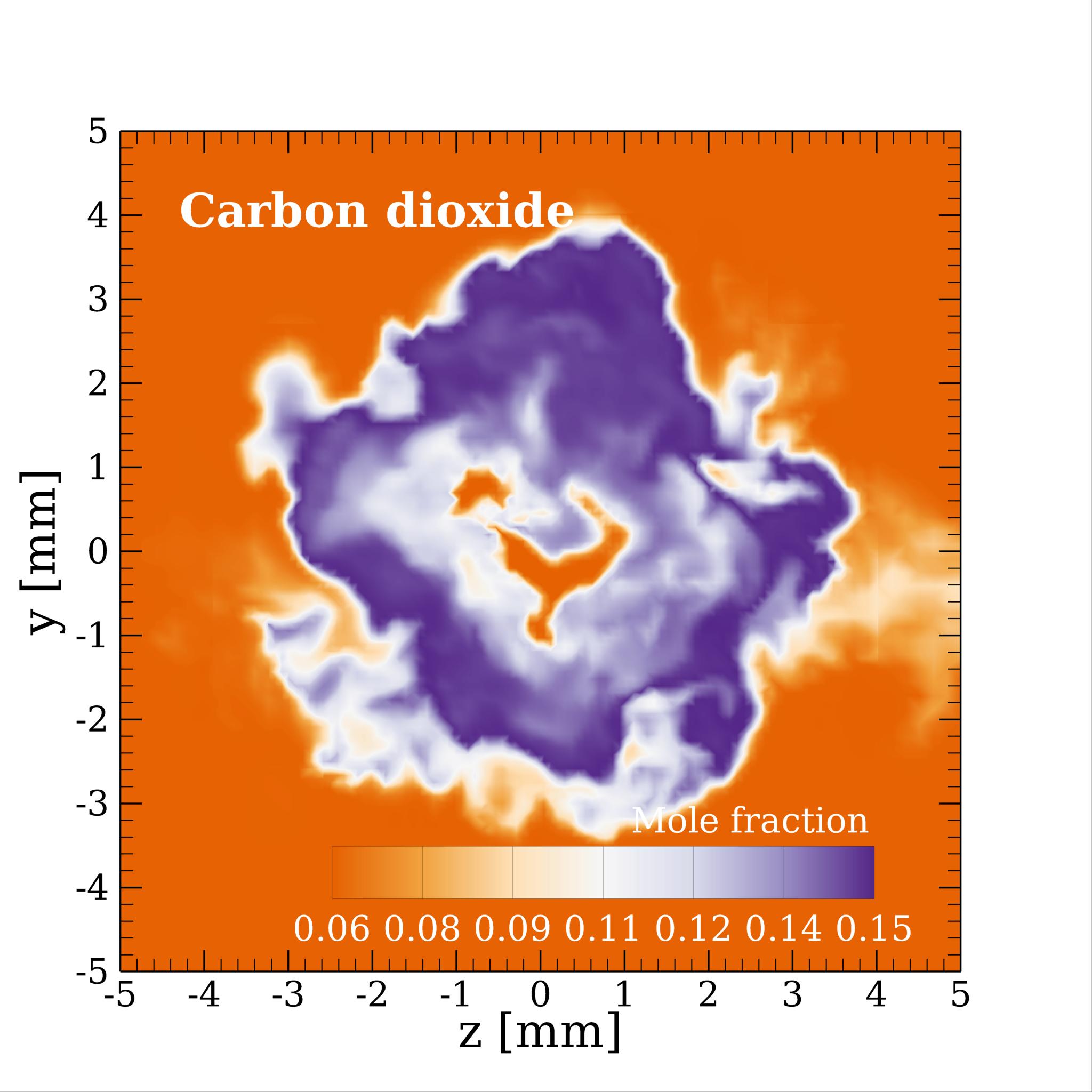}
         %\caption{$y=3sinx$}
         %\label{fig:three sin x}
     \end{subfigure}
     %%%%%%%%%%%%%%%%%%%%%%%%%%%%
     \begin{subfigure}[b]{0.25\linewidth}
         \centering
         \includegraphics[width=\linewidth]{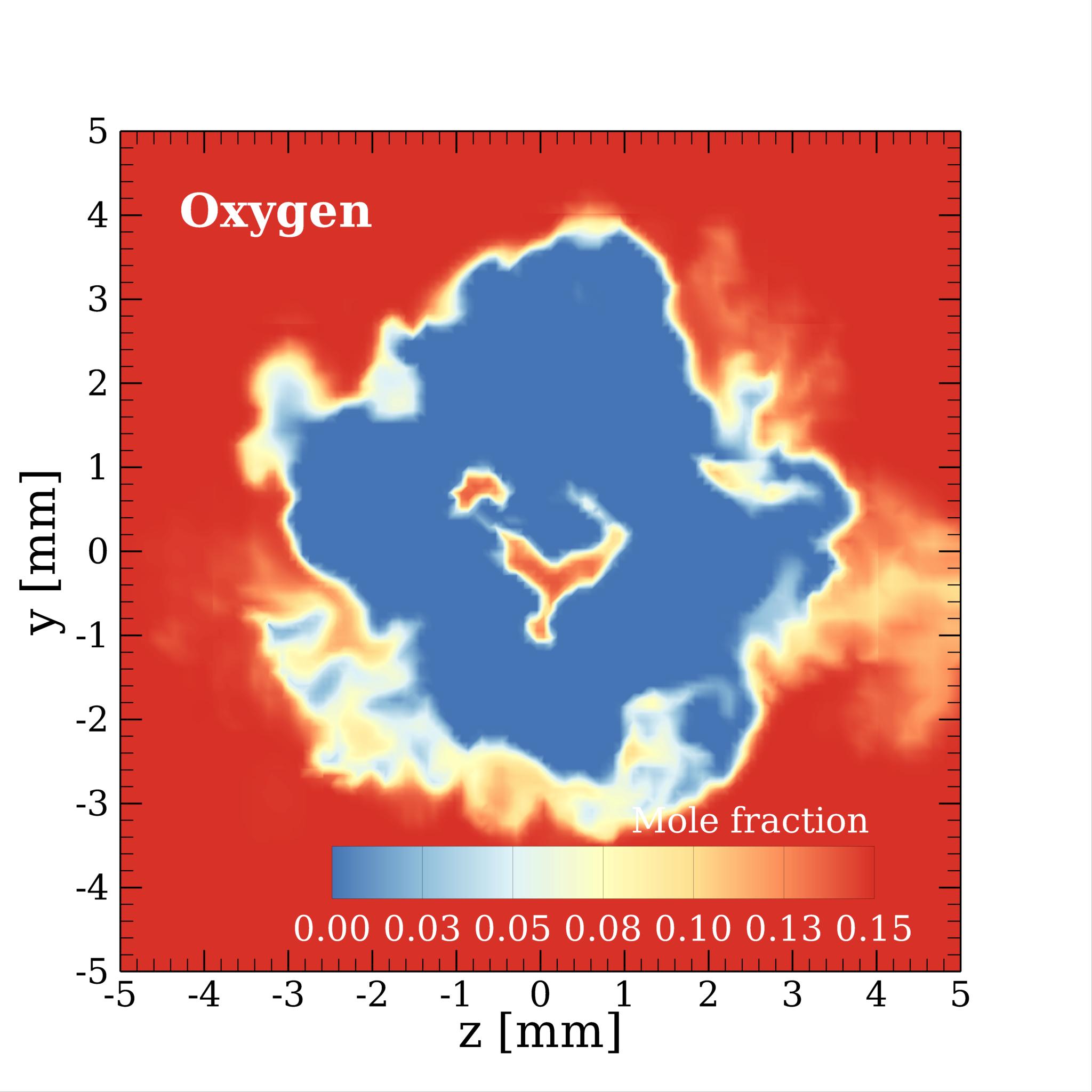}
         %\caption{$y=x$}
         %\label{fig:y equals x}
     \end{subfigure}
     \begin{subfigure}[b]{0.25\linewidth}
         \centering
         \includegraphics[width=\linewidth]{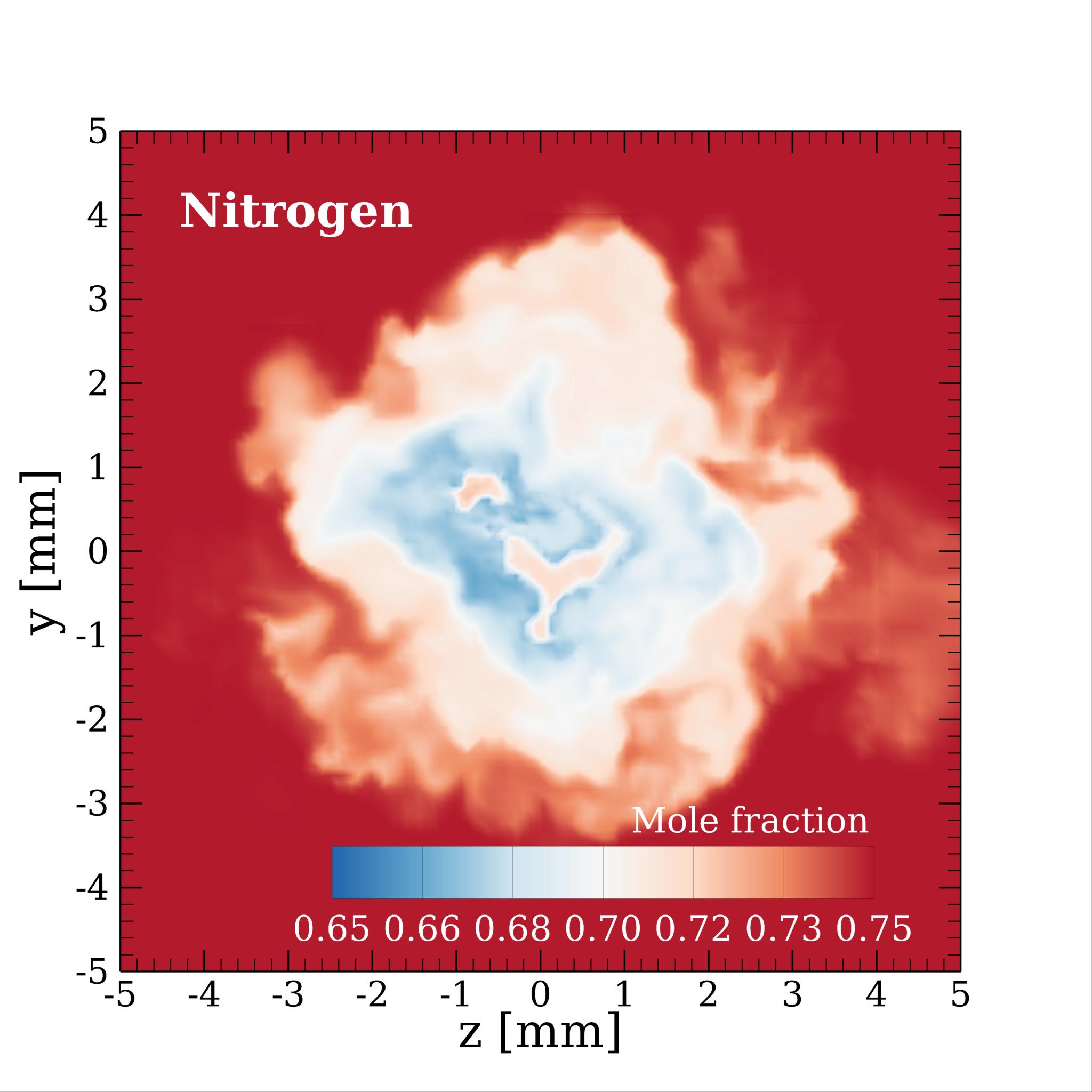}
         %\caption{$y=3sinx$}
         %\label{fig:three sin x}
     \end{subfigure}
     \begin{subfigure}[b]{0.25\linewidth}
         \centering
         \includegraphics[width=\linewidth]{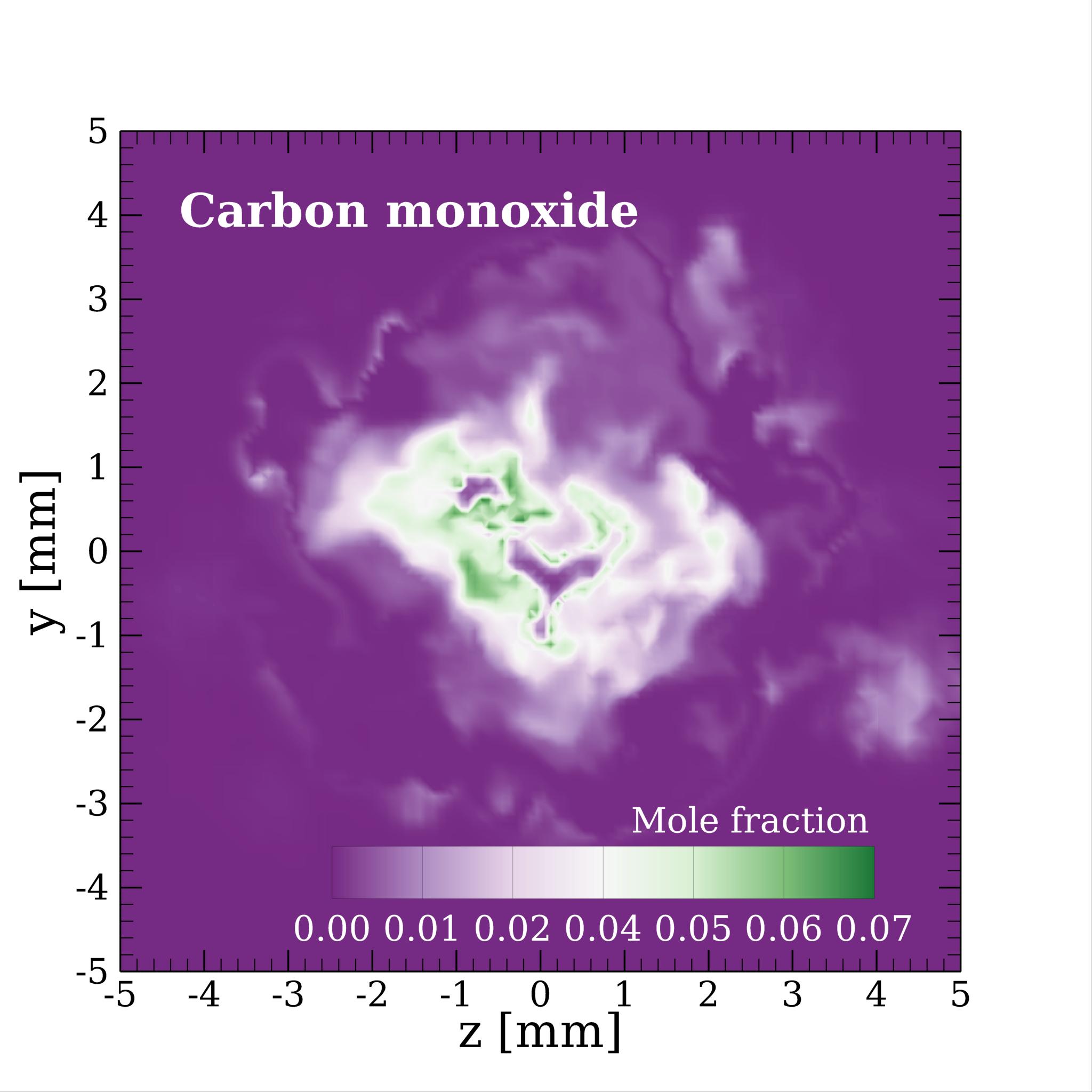}
         %\caption{$y=3sinx$}
         %\label{fig:three sin x}
     \end{subfigure}
     \caption{Instantaneous snapshots of species mole fractions located at plane $x=\SI{18}{\milli\meter}$ at \SI{590}{\micro\second} ASOI in the reacting case.}\label{fig:reacting_composition}
\end{figure}

Figure~\ref{fig:reacting_composition} shows species mole fraction contours on a plane normal to the axial direction at $x=\SI{18}{\milli\meter}$ and $t=\SI{590}{\micro\second}$, that is, at distance and time greater than LOL and IDT.  Accordingly, there exists no liquid phase at this location. The molar composition indicates partially premixed conditions of the fuel and environment even in the core of the reacting jet. The molar composition at the core is more close to the saturated vapor phase.

\begin{figure}[!htb]
         \centering
     \begin{subfigure}[b]{0.31\linewidth}
         \centering
         \includegraphics[width=\linewidth]{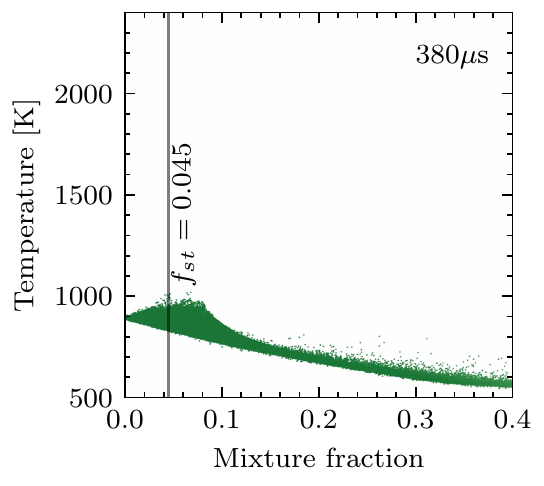}
         %\caption{$y=x$}
         %\label{fig:y equals x}
     \end{subfigure}
     \begin{subfigure}[b]{0.31\linewidth}
         \centering
         \includegraphics[width=\linewidth]{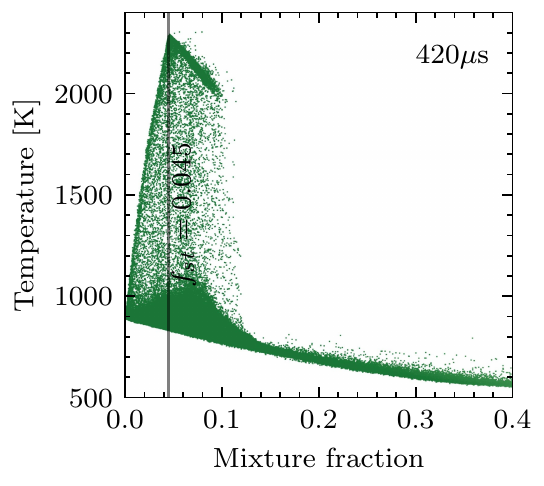}
         %\caption{$y=3sinx$}
         %\label{fig:three sin x}
     \end{subfigure}
     \begin{subfigure}[b]{0.31\linewidth}
         \centering
         \includegraphics[width=\linewidth]{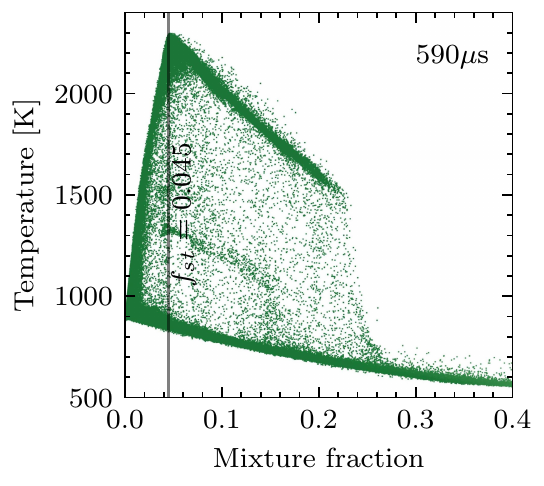}
         %\caption{$y=3sinx$}
         %\label{fig:three sin x}
     \end{subfigure}
     \caption{Temporal evolution of the temperature in mixture fraction space close and after the ignition time.}\label{fig:reacting_temperatureMixturefraction}
\end{figure}

The auto-ignition process is illustrated in Fig.~\ref{fig:reacting_temperatureMixturefraction}, where we show global scatter plots of the temperature in mixture fraction space at three different time instants. The first snapshot taken at \SI{380}{\micro\second} is representative for the very late pre-ignition state, where significant low-temperature reactions are occurring and temperature is slowly increasing around the stoichiometric mixture fraction. The spatial distribution of these low-temperature kernels is shown in Fig.~\ref{fig:reacting_snapshots}. The second snapshot is taken at \SI{420}{\micro\second}, that is, shortly after the auto-ignition time ($t \approx \SI{400}{\micro\second}$). Several high-temperature flame kernels have been formed in regions with a stoichiometric mixture fraction, where enough radicals and intermediate species coexist. In the third snapshot at \SI{590}{\micro\second}, the ignition process is completed and has provided enough thermal energy for the propagation of the flame. We see that the high-temperature reactions are spreading into the fuel-rich regime.

\section{Discussion}
\begin{figure}[!htb]
 \centering
\includegraphics[width=4.24in]{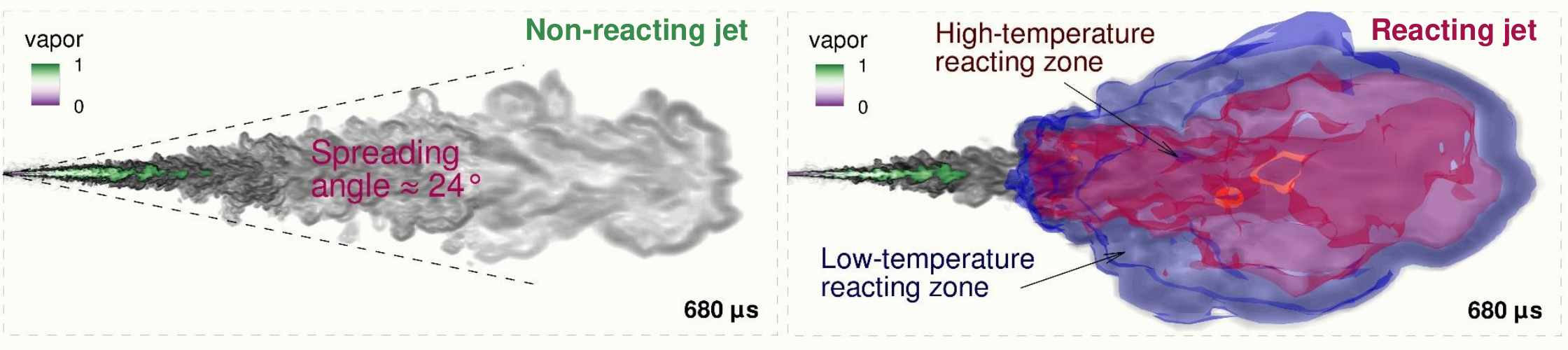}
\caption{LES data of the transcritical injection of n-dodecane (C$_{12}$H$_{26}$ at \SI{363}{\kelvin}) into a preheated chamber at \SI{900}{\kelvin} with and without oxygen. Operating pressure is \SI{60}{\bar} which exceeds the critical pressure of n-dodecane.}\label{fig:spraya}
\end{figure}

The above presented results of ECN Spray-A simulations demonstrate the potential of LES-MT modelling for the accurate prediction of reacting and non-reacting turbulent multiphase fluid flows at transcritical pressures. The proposed LES-MT method is free from the main limitations of classical Lagrangian methods, which do not account for the solubility of real-fluid mixtures at elevated pressures and additionally may suffer from inaccurate density predictions if ideal-gas models are used, and Eulerian dense-gas (DG) models, which can become arbitrarily inaccurate in the two-phase region (where they predict negative pressures, e.g.). Moreover, high-pressure combustion typically initiates in low-temperature reacting zones in which the fluid's state and transport properties deviate strongly from ideal-gas laws. Making ideal-gas assumptions, which is very common in combustion simulations, or the assumption of a dense-gas without accounting for local two-phase regions, can lead to large errors and uncertainties in transcritical combustion simulations. Furthermore, the LES-MT method provides the possibility of applying different reaction mechanism in the liquid and vapor phases, whereas classical DG methods have to apply the reaction mechanism on the overall composition, including possible condensates. For these reasons, we emphasise that real-fluid and phase-equilibrium effects should be considered in transcritial jets and flames. Doing so can yield very accurate predictions for the spreading angle, liquid penetration lengths, vapor penetration lengths, ignition delay time, and flame lift-off length, such as shown in Fig.~\ref{fig:spraya} and discussed in the previous sections. 

The major drawback of the LES-MT method might be the highly non-uniform computation costs, which limits the parallel scalability of flow solvers with domain decomposition methods. It should be also mentioned that we have applied LES in a rather straightforward way without specifically addressing turbulence-combustion interactions other than by providing reasonable spatial and temporal resolution. The unresolved thermodynamic microstructure has without doubt nonlinear effects on the resolved-scale solution, and it seems reasonable to assume that transcritical high-pressure flows are much stronger affected than ideal-gas flows. Hence, the quantification of uncertainties resulting from the unresolved microstructure on the resolved pressure, temperature, and subgrid-scale terms are subject of our future studies.

\section{Conclusions}

We have presented a real-fluid finite-rate chemistry multiphase thermodynamics model for the numerical simulation of transcritical vaporization and auto-ignition of a cold or cryogenic fuel injected into a hot high-pressure environment. The methodology is based on solving the fully conservative form of the compressible multi-species Navier-Stokes equations along with real-fluid kinetic and caloric state equations. These state equations provide accurate real-fluid thermodynamic properties for multi-component fluids that can exist either in a single phase or undergo phase transitions during vaporization and condensation. Multiphase and real-gas effects are also considered in the finite-rate chemistry model, which we propose to base on the fugacity of the species. 

The methodology has been demonstrated and validated for the transcritical reacting and non-reacting ECN Spray-A. LES results obtained with the proposed thermodynamics models are in excellent agreement with experimental reference data for the non-reacting and the reacting cases. The time evolution of temperature in the mixture fraction space proves the existence of the low-temperature and high-temperature combustion stages in the auto-ignition process of Spray-A that have been reported experimentally. 

The very good agreement of auto-ignition time, flame lift-off length, and flame structure might be surprising because only a simple two-step reaction mechanism has been applied. However, one should note that Hakim's mechanism has been specifically calibrated for the considered flow conditions including low- and high-temperature ignition stages. In our LES, the real-gas vapor-liquid equilibrium accurately determines the saturated vapor composition, and this composition and the rate with which it mixes with the oxidizer determines when and where auto-ignition takes place.

\medskip

\section*{Acknowledgements}
This paper is submitted as part of the themed issue  “Development and Validation of Models for Turbulent Reacting Flows” in honor of Professor Michael Pfitzner on the occasion of his retirement. The second author would like to thank Michael for motivating him to start working on this topic and for the very productive collaboration over many years.

\section*{Author Declarations}
\paragraph{Funding:}This work is funded by the Netherlands Organisation for Scientific Research (NWO) under Contract No.~680.91.082. Simulations have been performed on DelftBlue at the Delft High Performance Computing Centre (DHPC).
\paragraph{Competing Interests:}The authors have no conflicts to disclose.
\paragraph{Data:}The data that support the findings of this study are available from the corresponding author upon reasonable request.

\bibliographystyle{unsrt}
\bibliography{references}

\end{document}